\documentclass[prl,twocolumn,nopacs,nofootinbib,superscriptaddress]{revtex4-2}

\usepackage{graphicx}
\usepackage{dcolumn}

\usepackage{color}

\usepackage[dvips]{epsfig}
\usepackage{amssymb}

\usepackage{amsmath}

\usepackage{eucal}
\usepackage{mathrsfs}
\usepackage{amsmath}
\usepackage{amssymb}
\usepackage{amsfonts}
\usepackage{latexsym}
\usepackage{color}

\newcommand{\bs}{\boldsymbol}

\begin{document}

\title{Ultrafast reversible self-assembly of living tangled matter}
\author{Vishal P. Patil}
\thanks{V.P.P. and H.T. contributed equally to this work.}
\address{School of Humanities and Sciences, 
Stanford University,
450 Serra Mall, Stanford, CA 94305}

\author{Harry Tuazon}
\thanks{V.P.P. and H.T. contributed equally to this work.}
\address{School of Chemical and Biomolecular Engineering, Georgia Institute of Technology, Atlanta, GA 30318}

\author{Emily Kaufman}
\address{School of Chemical and Biomolecular Engineering, Georgia Institute of Technology, Atlanta, GA 30318}
\author{Tuhin Chakrabortty}
\address{School of Chemical and Biomolecular Engineering, Georgia Institute of Technology, Atlanta, GA 30318}

\author{David Qin}
\address{Wallace H. Coulter Department of Biomedical Engineering, Georgia Institute of Technology, Atlanta, GA 30332}

\author{J\"orn Dunkel}
\thanks{Correspondence to dunkel@mit.edu; saadb@chbe.gatech.edu}
\address{Department of Mathematics, 
Massachusetts Institute of Technology, 
77 Massachusetts Avenue, 
Cambridge, MA 02139}

\author{M. Saad Bhamla}
\thanks{Correspondence to dunkel@mit.edu; saadb@chbe.gatech.edu}
\address{School of Chemical and Biomolecular Engineering, Georgia Institute of Technology, Atlanta, GA 30318} 

\date{\today}

\begin{abstract}
Tangled active filaments are ubiquitous in nature, from chromosomal DNA and cilia carpets to root networks and worm blobs. How activity and elasticity facilitate collective  topological transformations in living tangled matter is not well understood. Here, we report an experimental and theoretical study  of  California blackworms (\textit{Lumbriculus variegatus}), which  slowly form tangles over minutes but can untangle in milliseconds. Combining ultrasound imaging, theoretical analysis and simulations, we develop and validate a mechanistic model that explains how the kinematics of individual active filaments determines their emergent collective  topological dynamics. The model reveals that resonantly alternating helical waves enable both tangle formation and ultrafast untangling. By identifying generic dynamical principles of topological self-transformations, our results can provide guidance for  designing new classes of   topologically tunable  active  materials.

\end{abstract}
\maketitle

Filaments and fibers are a crucial building block of complex matter, giving rise to a broad variety of morphologies with distinct mechanical and topological properties. From entangled polymeric systems~\cite{mirny2011fractal, gardel2004elastic, peterson2021constitutive, lua2006statistics, orlandini2019synergy,grosberg1996flory, duprat2012wetting, winkler2020physics}, active cilia carpets~\cite{gilpin2020multiscale} and worms blobs~\cite{ozkan2021collective}, to everyday macroscopic materials including yarn~\cite{warren2018clothes, chopin2022tensional}, hair~\cite{goldstein2012shape} and fabrics~\cite{hu2004structure}, the propensity of the underlying filaments to tangle~\cite{belmonte2001dynamic,soh2019self, huang2021numerical} is responsible for the emergent dynamics of a range of biological and physical systems. The resulting topological obstructions~\cite{panagiotou2010linking, panagiotou2021vassiliev, gonzalez1999global} induce constraints on motion that can lead to materials with different transport~\cite{heeremans2022chromatographic, liu2021isotopy}, stress-response~\cite{panagiotou2019topological, becker2022active, sano2022exploring, johanns2021shapes} and energetic properties~\cite{liu2021isotopy}. Although tangling can inhibit functionality in common materials~\cite{raymer2007spontaneous}, the topological control and manipulation of amorphous tangles has remained a tantalizing theoretical and experimental challenge. In particular, the question of how to quickly unravel a complex tangle presents a historically famous problem~\cite{kauffman2012hard, ferschweiler2022percolation}, of equal importance to comb makers~\cite{plumb2022combing} and  coiffeurs~\cite{goldstein2012shape, hair_video} as to cells~\cite{sato2021multistep} and  crawling animals~\cite{drewes1989hindsight,drewes1999helical}.

Living tangled matter, consisting of filamentary objects which can braid and wind around each other, represents an important class of topological active matter~\cite{drewes1999helical, hu2016entangled, shankar2022topological}. Such systems are often capable of dynamically controlling their topological state,  and exploiting apparently disordered tangles as a resource~\cite{drewes1999helical,ozkan2021collective, nguyen2021emergent, zirbes2012self, franks2016social}. A particularly striking example is the California blackworm (\textit{Lumbriculus variegatus})~\cite{drewes1999helical}, owing to its ability to assemble into three-dimensional (3D)  tangles over the course of minutes, and rapidly disentangle in milliseconds (movie S1). Biologically, a blackworm collective uses the tangled state to efficiently execute a range of essential functions, such as temperature maintenance and moisture retention~\cite{tuazon2022oxygenation,ozkan2021collective, franks2016social}. Perhaps even more importantly, the ability to rapidly escape from the tangle is an important predation response~\cite{drewes1989hindsight}. The biophysical mechanisms by which basic filamentous organisms can achieve such ultrafast untangling have remained unknown. Here, motivated by this question, we combine ultrasound imaging experiments and elasticity theory to explain how individual worm gaits lead to large-scale topological dynamics and transitions. By mapping worm tangling to percolation problems and picture-hanging puzzles~\cite{demaine2014picture}, our analysis shows how resonantly tuned helical waves enable self-assembly and rapid unknotting of tangled matter, thus revealing a generic dynamical principle that can guide the design of novel active materials.

\begin{figure*}[t]
	\centering
	\includegraphics[width=\textwidth]{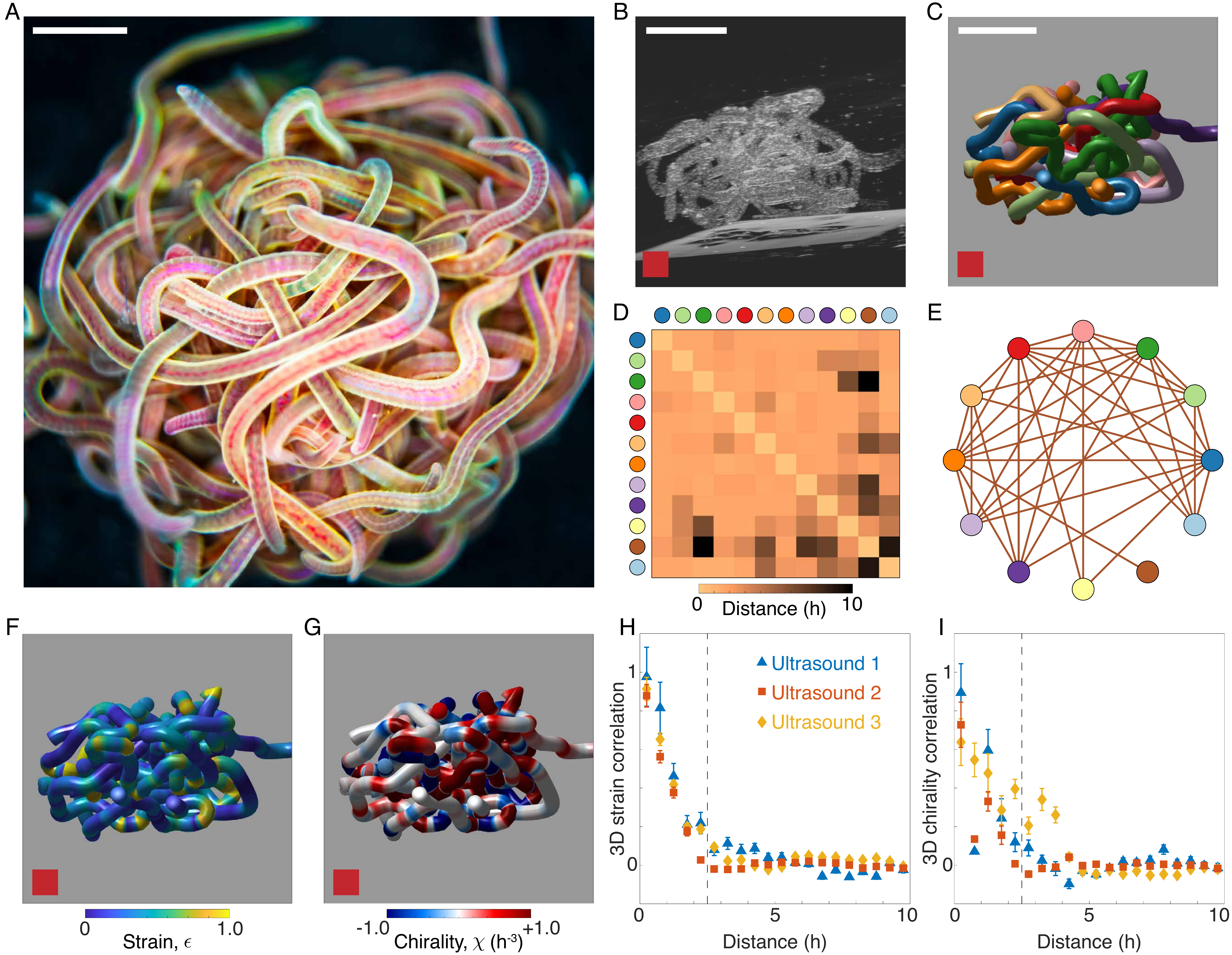}
	\caption{\textbf{3D ultrasound data reveal the mechanical structure of active, biological worm tangles}
	(A)~Topologically complex tangle formed by \textit{Lumbriculus variegatus} consisting of approximately 200 worms. Scale bar 3mm.
	(B,C)~Ultrasound imaging reveals the interior structure of a 12-worm tangle. Scale bar 5mm.
	(D,E)~The contact matrix and contact graph confirm that the worm tangle is a strongly interacting system.
	(F,G)~3D experimental data enable the visualization of strain $\epsilon$, and chirality $\chi$, fields within the tangle, revealing that the worms form achiral tangles.
	(H,I)~Decorrelation of strain, $\rho_C(\epsilon(\bs{x}), \epsilon(\bs{y}))$, and chirality, $\rho_C(\chi(\bs{x}), \chi(\bs{y}))$, over distances of $|\bs{x} - \bs{y}| \approx 2.5h$ (dotted lines) demonstrates the limits of a continuum elastic theory for worm tangles. The decorrelation length scale indicates the existence of an effective radius, $h_{\text{eff}} \sim 1.25h$, arising from the preparation of tangles for ultrasound (Methods).
	}
	\label{fig1}
\end{figure*}

\begin{figure*}[t]
	\centering
	\includegraphics[width=\textwidth]{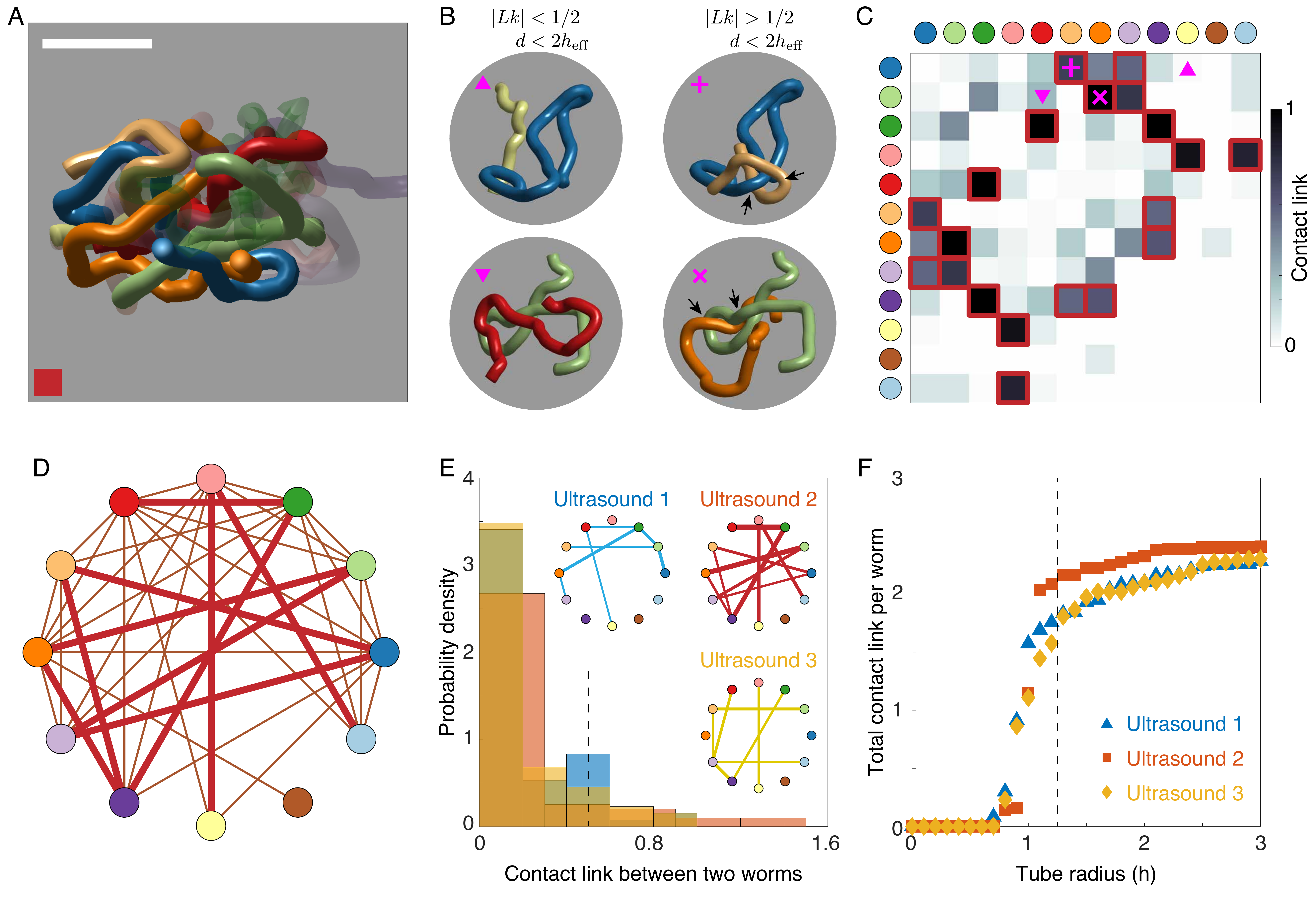}
	\caption{\textbf{Topological structure of worm tangles}
	(A)~3D ultrasound reconstructions (as in Fig.~1B,C) allow for the individual topological interactions between chosen worms (solid color) to be mapped in detail. Scale bar 5mm.
	(B)~Topological analysis enables the classification of tangle structure by distinguishing between contact (left column) and braiding interactions (right column), which are defined by having linking number $|Lk|>1/2$.
	(C)~Contact link, $cLk$, defined as the absolute value of the link between worms separated by at most $2h_{\text{eff}}$, identifies the strongest topological interactions within the tangle. The contact link between non-touching worms is 0. Pairs of worms with $cLk>1/2$ are highlighted in red.
	(D)~The tangle graph provides a sparser representation of tangle state than the contact graph. Edges are present between pairs of worms with $cLk>1/2$, i.e. worms that both touch and have $|Lk|>1/2$ (red bordered squares in C).
	(E)~The probability distribution of the contact link between two worms is stable across ultrasound data sets. Pairs of worms with contact link greater than $1/2$ (dotted line) lead to edges in the corresponding tangle graphs (inset), with edge thickness given by the value of the contact link.
	(F)~Increasing the tube radius of the worm curves modifies the contact structure of the tangle and thus increases the total contact link (SI). The radius dependence of total contact link is similar across different tangles, and indicates the presence an effective radius as in Fig.~\ref{fig1}H,I, that is distinct from the true radius, $h$.
	}
	\label{fig2}
\end{figure*}

\begin{figure*}[t!]
	\centering
	\includegraphics[width=\textwidth]{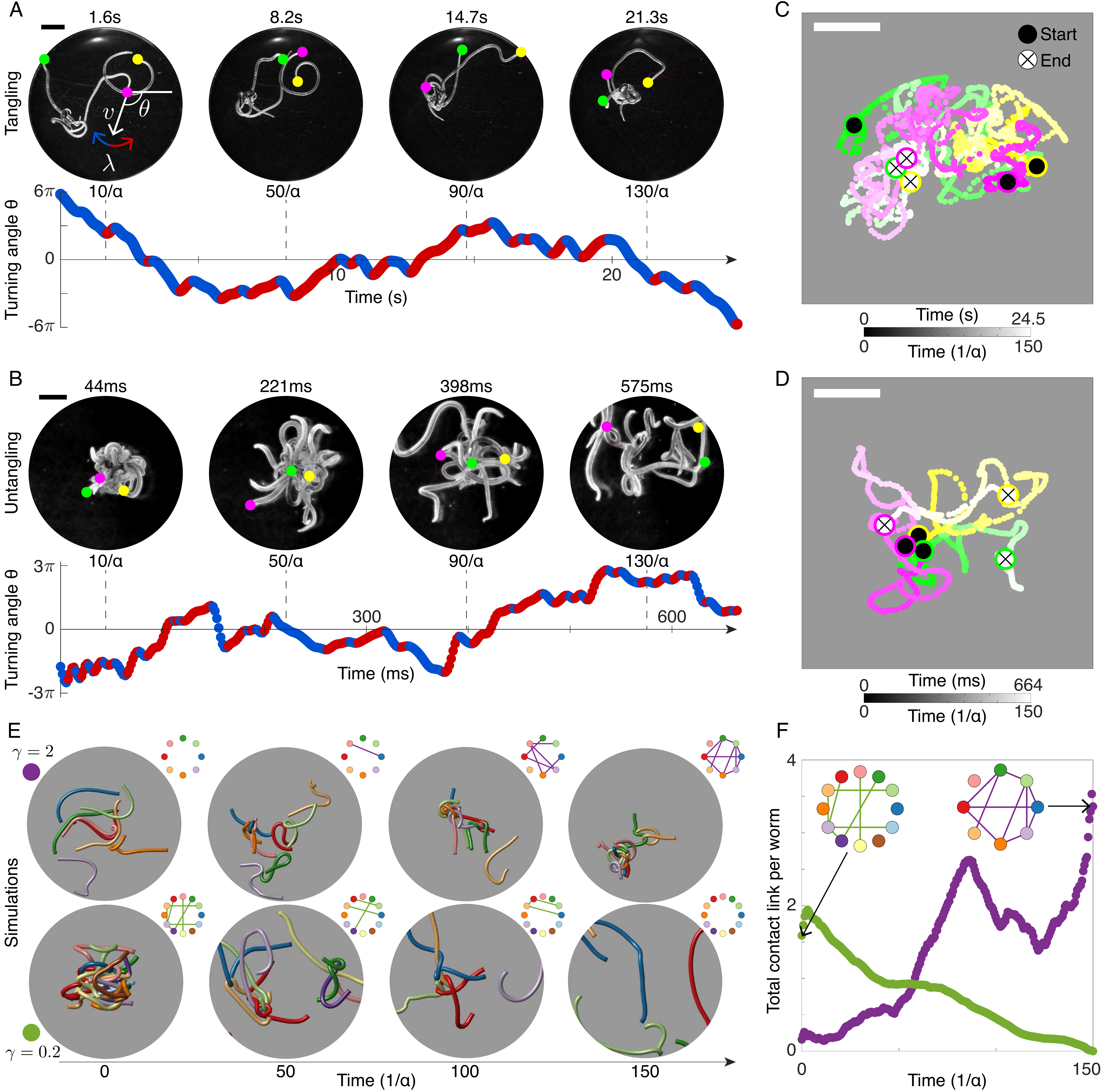}
	\caption{\textbf{Resonant helical worm head dynamics give rise to numerically reproducible weaving and unweaving gaits.}
(A,B)~Experimentally observed worm head trajectories~\cite{schneider2012nih,tinevez2017trackmate} projected into 2D can be approximated by their angular direction, $\theta(t) = \arg \dot{\bs{x}}(t) $, in both the tangling (A) and untangling (B) cases (movie S2). $\theta$ is characterized by an average turning rate, $\alpha =\langle|\dot{\theta}|\rangle$, and a rate of switching from left turning (red points, $\dot{\theta} >0$) to right turning (blue points, $\dot{\theta} < 0$). The chirality number, $\gamma = \alpha/2\pi\lambda$, captures the difference between weaving ($\gamma=0.68$) and unweaving ($\gamma=0.36$) gaits. $\alpha^{-1}$ defines an intrinsic timescale for tangle assembly and disassembly. Scale bars 3mm.
(C,D)~Experimentally measured head trajectories of 3 worms (different colors) executing the tangling (C) and untangling (D) gaits demonstrate the formation (C) or removal (D) of topological obstructions within a similar time in units of $\alpha^{-1}$. Scale bars 5mm.
(E)~Simulations of active Kirchhoff filaments demonstrate that the gaits described in (A,B) are sufficient for reversible tangle self-assembly (movie S2). The topological state is quantified using tangle graphs (inset). Tangling filaments have large $\gamma$ (top row E and A) and untangling filaments have small $\gamma$ (bottom row E and B). The initial tangled state (bottom row E) is obtained from 3D ultrasound reconstruction. Average worm lengths range from $40\,$mm (top row) to $28\,$mm (bottom row), with radius $0.5\,$mm throughout. Displayed worms are thickened to aid visualization.
(F)~The total contact link per worm (Fig.~\ref{fig2}) obtained from simulations reveals the rate at which tangles form (purple dots, top row panel E) and unravel (green dots, bottom row panel E).
	}
	\label{fig3}
\end{figure*}

Blackworms are capable of assembling into topologically intricate tangles consisting of anywhere from 5 to 50000 worms~\cite{ozkan2021collective} (Fig.~\ref{fig1}A). Our ultrasound experiments, conducted on worm tangles immobilized in gelatin, allow for the reconstruction of the 3D  structure of a living tangle (Fig.~\ref{fig1}B,C; Methods). This reveals a picture of the tangle as a strongly interacting system, in which the worms are tightly packed (Fig.~\ref{fig1}D), and most worms are in contact with most other worms (Fig.~\ref{fig1}E). In addition to the arrangement of contact, the non-topological structure of the worm tangle can also be described by the variation of geometric quantities both within and between different worms. To analyze the tangle geometry, we approximate each worm as a curve $\bs{x}(s)$, parameterized by arc length $s$, which can be characterized by local in-plane curvature, $\kappa(s)$, and an out-of-plane 3D torsion, $\tau(s)$ (SI). These give rise to bending strain, $\epsilon = \kappa h$ (Fig.~\ref{fig1}F), and chirality, $\chi = \kappa^2\tau$ (Fig.~\ref{fig1}G), where $h$ is the worm radius. The 3D distribution of both strain and chirality is primarily heterogeneous (Fig.~\ref{fig1}F,G). More formally, the correlation coefficients for strain and chirality, $\rho_C(\epsilon(\bs{x}), \epsilon(\bs{y}))$ and $\rho_C(\chi(\bs{x}), \chi(\bs{y}))$, decay rapidly as functions of the spatial separation, $|\bs{x} - \bs{y}|$ (Fig.~\ref{fig1}H,I). For small values of $|\bs{x} - \bs{y}|$, the correlation functions are dominated by intraworm interactions, but decorrelation occurs once $\rho_C$ begins to include interworm effects. In particular, $\rho \approx 0$ for both strain and chirality once $|\bs{x} - \bs{y}| > 2.5h$, which indicates the existence of an effective radius, $h_{\text{eff}} = 1.25 h$. This effective radius is a signature of the ultrasound protocol (Methods), which requires the tangles to undergo a small dilation. The rapid decorrelation demonstrates that strain and chirality are not described by 3D continuum fields, illustrating the difficulty of constructing a continuum theory for the living tangle. Understanding the mesoscale structure of the tangle requires moving beyond purely geometrical properties.

Topological analysis of the tangle geometry allows us to distinguish between different forms of contact. The intuitive notion that worms which braid should interact more strongly than worms which simply touch can be captured by considering the linking number~\cite{qu2021fast}, $Lk$, of the $i$'th worm and the $j$'th worm
\begin{align}
    Lk_{ij} = \frac{1}{4\pi} \int ds d\sigma \; \Gamma_{ij} \cdot \left( \partial_s \Gamma_{ij} \times \partial_{\sigma} \Gamma_{ij} \right) 
\end{align}
where $\Gamma_{ij}(s, \sigma) = (\bs{x}_i(s) - \bs{x}_j(\sigma) ) /|\bs{x}_i(s) - \bs{x}_j(\sigma)|$, and $\bs{x}_i, \bs{x}_j$ are the curves representing the $i$'th and $j$'th worms. Although traditionally defined only for closed curves, the linking number of open curves quantifies entanglement by taking an average of the amount of braiding in every 2D projection~\cite{qinami2017quantifying} (SI). Visually, pairs of worms with $|Lk| > 1/2$ appear to wind around each other (Fig.~\ref{fig2}A,B). However, $Lk$ is not sensitive to contact, which must ultimately mediate every worm-worm interaction. Accordingly, we define a more sensitive measure, termed contact link $cLk$, by setting $cLk = |Lk|$ for worms in contact, and $cLk=0$ otherwise. In contrast to the contact matrix (Fig.~\ref{fig1}D), the contact link matrix (Fig.~\ref{fig2}C) identifies a far smaller number of key interactions, thus providing a sparser representation of tangle state. This is particularly evident from the tangle graph (Fig.~\ref{fig2}D), which shows worm-worm interactions with $cLk>1/2$. The robustness of contact link as a measure of tangling is evident through its behavior across different ultrasound data sets. For example, the probability distribution of the contact link between two worms, a measure of topological interaction strength, retains a characteristic shape for all three tangles (Fig.~\ref{fig2}E). Additionally, the total contact link (SI), obtained by summing all the pair contact links from Fig.~\ref{fig2}C, is sensitive to the contact structure of the tangle. In particular, the total contact link as a function of worm radius (Fig.~\ref{fig2}F) behaves similarly as the worms are thickened from zero radius to larger radii. Thus, by incorporating topological information~\cite{qinami2017quantifying, panagiotou2013quantifying} as well as geometric information, contact link $cLk$  captures core structural motifs that are reproducible over different experiments. In particular $cLk$ will enable us to compare experimentally observed worm tangles with tangled structures generated by dynamical simulations.

The ability of the blackworm to form tangles over minutes (Fig.~\ref{fig3}A), but rapidly unravel in milliseconds (Fig.~\ref{fig3}B) is a key biological and topological puzzle~\cite{zirbes2012self, nguyen2021emergent}. To understand the dynamical process that gives rise to tangle formation, we experimentally studied the head trajectories of single worms (Fig.~\ref{fig3}A-D; Methods). Since these experiments were performed in a shallow fluid well (height $\sim2\,$mm), the projection of the trajectories into 2D (Fig.~\ref{fig3}A-D) does not cause significant information loss. To capture the winding motions associated with braiding and unbraiding, we assume the worm head has preferred speed $v = \langle|\dot{\bs{x}}(t)|\rangle$, and focus on the worm turning direction, $\theta(t) = \arg \dot{\bs{x}}(t)$. The $\theta$ trajectories can be approximately described in terms of two parameters, the average angular speed $\alpha = \langle|\dot{\theta}|\rangle$ (Fig.~\ref{fig3}A,B) and the rate $\lambda$ at which $\dot{\theta}$ changes sign. These quantities can be estimated from the noisy trajectory data (SI). Although the characteristic timescales $\alpha^{-1}$ for slow tangling and ultrafast untangling differ by 2 orders of magnitude, rescaling the $\theta$ trajectories for each gait by  $\alpha^{-1}$ reveals a similar underlying dynamics  (Fig.~\ref{fig3}A,B). This similarity reflects the fact that locomotion machinery is biologically constrained~\cite{hu2009mechanics}, and indicates that tangling and untangling can be captured by the same mathematical model. To confirm this, we first formulate a minimal 2D model of worm head dynamics which we will then generalize to a full 3D dynamical picture.

A minimal 2D model can be constructed by focusing on the helical worm head dynamics identified experimentally (Fig.~\ref{fig3}). In particular, the quantities $\alpha, \lambda$ and $v$ discussed above motivate the following stochastic differential equation (SDE) model for a worm-head trajectory
\begin{align}\label{SDE}
    \dot{\bs{x}} = v \bs{n}_\theta + \bs{\xi}_T , \qquad \dot{\theta} = \sigma(t; \lambda) \alpha + \xi_R
\end{align}
where $\bs{\xi}_T, \xi_R$ are noise terms, $\bs{n}_\theta$ is a unit vector in the $\theta$ direction and $\sigma(t;\lambda)$ switches between $+1$ and $-1$ at rate $\lambda$ (SI). These trajectories can be further classified by dimensionless parameters. In particular, the chirality number, $\gamma = \alpha/2\pi\lambda$, distinguishes between the tangling and untangling gaits (Fig.~\ref{fig3}A,B). This non-dimensional parameter corresponds to the average number of right-handed or left-handed loops traced out by the worm before changing direction and provides an intuitive way of understanding the topological properties of each gait. When $\gamma$ is large, worms wind around each other before switching direction, thus producing a coherent tangle. On the other hand, for small $\gamma$, the worms change direction before they are able to wind around one another and so remain untangled. Our trajectory model thus explains how the characteristic helical waves produced by untangling worms mediate topology (movie S2).

We next show that these conclusions generalize to a full 3D mechanical model of worm gaits. To model the worms, we performed elastic fiber simulations where the worms are treated as Kirchhoff filaments~\cite{bergou2008discrete,bergou2010discrete,audoly2010elasticity, patil2020topological, patil2020discharging,wolgemuth2000twirling, tong2022fully, gilpin2015worms} with active head dynamics (SI). The head motions are prescribed by the SDE model \eqref{SDE} together with additional 3D drift (SI). The resulting worm collectives can form 3D tangled structures (Fig.~\ref{fig3}E) consistent with those seen in our experiments, as quantified by contact link (Fig.~\ref{fig3}F). In particular, the tangling and untangling behavior in these simulations appears to be a function of the chirality number, $\gamma$, further confirming its importance (Fig.~\ref{fig3}E,F; movie S2). This formulation of a 3D dynamical model allows us to understand how the dynamics of single worms produces worm collectives with distinct topologies.

\begin{figure*}[t]
	\centering
	\includegraphics[width=\textwidth]{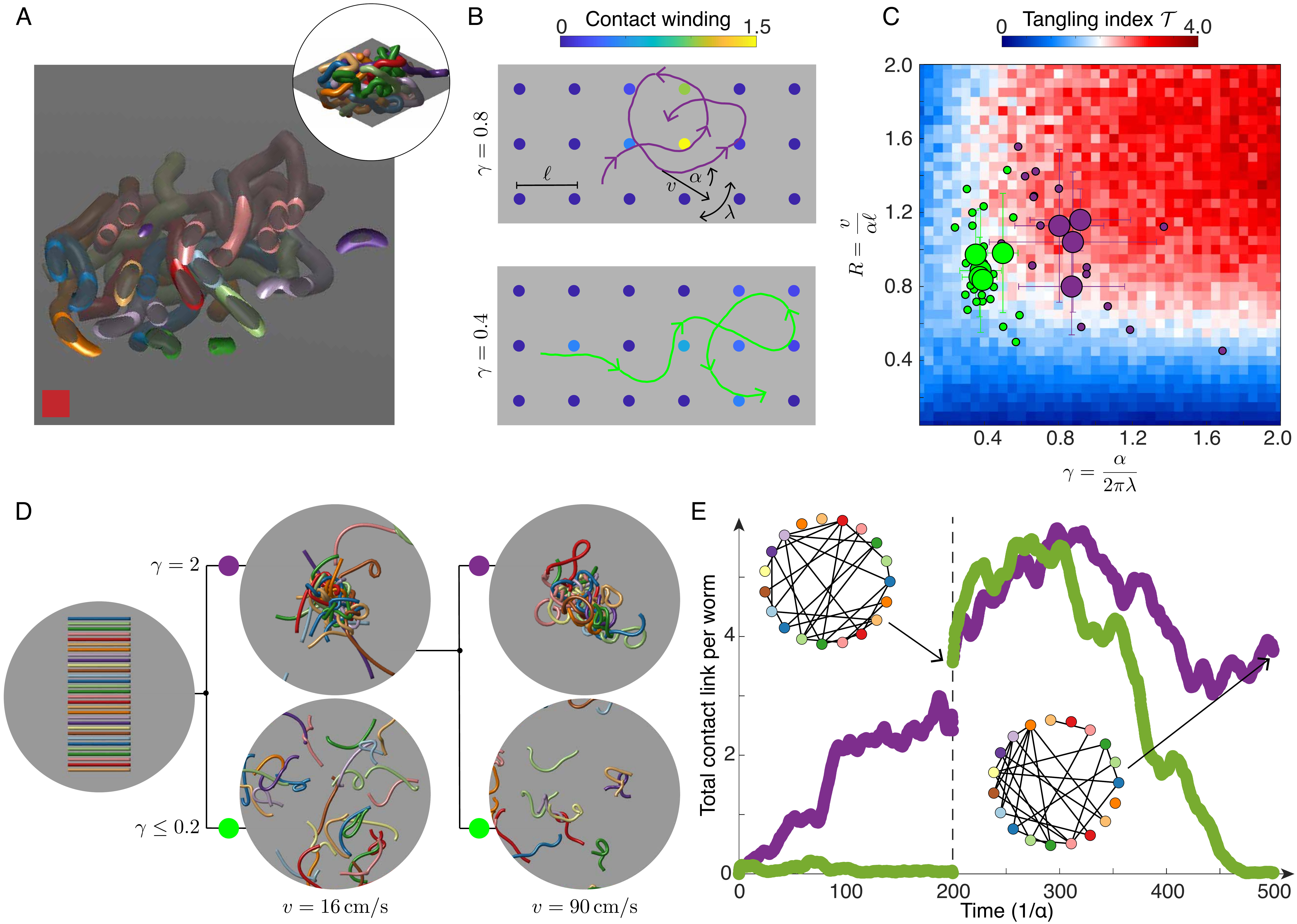}
	\caption{\textbf{Bio-inspired tangling model reveals phase diagram underlying topological assembly and manipulation of generic tangles.}
	(A)~2D cross sections of 3D ultrasound reconstructions indicate the obstacle landscape faced by a worm exhibiting quasi-2D motion.
	(B)~A 2D mean-field tangling model measures the winding of a worm head trajectory (purple and green curves) around fixed obstacles in the plane (filled circles). Contact winding, $cW_p$, around obstacles which are greater than $\ell/3$ from the trajectory is $0$ (SI). Points with $cW_p>1$ contribute to the tangling index, $\mathcal{T}$, of a trajectory  (eq.~\eqref{tangling_index}). Trajectories with large chirality number~$\gamma$ change direction frequently to minimize overall contact winding (bottom row). $R = v/\alpha\ell = 0.9$ for both trajectories.
	(C)~Measured values of $\gamma$ and $R$ for blackworms undergoing tangling (purple disks) or untangling (green disks) dynamics lie in regions of the tangle phase space corresponding to tangling (red, $\mathcal{T} > 2$) and untangling (blue, $\mathcal{T}< 2$), where the critical value $\mathcal{T}^\ast=2$ corresponds to a connected tangle graph, and hence a minimally tangled state. The untangling data consists of $N=25$ worms (small green disks) from $N=5$ separate untangling experiments, and the tangling data consists of $N=18$ worms (small purple disks) from $N=4$ separate tangling experiments. The large disks show mean values of $\gamma$ and $R$ obtained by averaging over all worms in a given experiment (Methods).
	(D)~Worm gaits predicted by the tangling phase diagram enable robust control of topological transitions (movie S3). Tangle formation and avoidance can be controlled at fixed $R$ by varying $\gamma$, both for low worm speeds $v$, (middle column, $30$ worms, $R=3.4$) and high worm speeds (right hand column, $19$ worms, $R=1.0$). Worms have length $40\,$mm and radius $0.5\,$mm. Displayed worms are thickened to aid visualization.
	(E)~Timescales of tangling and untangling from simulations in (D) are set by $\alpha^{-1}$, which varies from the low~$v$ simulations ($0<t<200/\alpha , \; \alpha^{-1} \approx 0.1\,$s) to the high~$v$ simulations ($t>200/\alpha, \; \alpha^{-1} \approx 4\,$ms). The largest cluster of touching worms produced by the low~$v$, large~$\gamma$ simulation is used as the initial condition for the high speed simulations (SI), causing an apparent jump in total contact link per worm at $t=200/\alpha$.  
	Tangle graphs (inset) illustrate the topological structure of the artificial tangles produced.
	}
	\label{fig4}
\end{figure*}

Based on the above analysis of the worm trajectories we can build a mean-field tangling model, which establishes a mapping between tangling and percolation (Fig.~\ref{fig4}). To formulate an analytically tractable model, we treat the worm motion as approximately 2D, so each worm effectively moves in a 2D slice of the 3D tangle (Fig.~\ref{fig4}A,B). As a given worm moves in a plane $P$, its head traces out a curve, $\bs{x}(t)$ (Fig.~\ref{fig4}B, purple and green curves), described by equation \eqref{SDE}. The worm can encounter obstacles, which represent intersections of the other worms with the plane $P$ (Fig.~\ref{fig4}B, colored circles). For simplicity, we treat these obstacles as forming a square lattice $\Lambda \subset P$, with spacing $\ell$ (Fig.~\ref{fig4}B), which corresponds to the tightness of the worm tangle (SI). The 3D notion of contact link between worms can be mapped to this 2D picture~\cite{demaine2014picture} by considering the winding of the trajectory $\bs{x}(t)$ around the obstacles $p\in\Lambda$. In particular, let $W_p$ be the winding number of $\bs{x}(t)$ around $p\in\Lambda$ after time $t=L/v$, the time taken for the worm-head to move one worm length, $L$ (SI). The contact winding, $cW_p$ of $\bs{x}(t)$ around $p$ is then $|W_p|$ if $\bs{x}(t)$ gets within a threshold distance of $p$ (SI), and is 0 otherwise (Fig.~\ref{fig4}B). As observed in experiments (Fig.~\ref{fig3}), topological properties such as the contact winding are sensitive to the chirality number $\gamma$ (Fig.~\ref{fig4}B). Thresholding and averaging all the contact winding numbers yields a tangling index
\begin{align}\label{tangling_index}
\mathcal{T} = \left\langle \sum_{p \in \Lambda} \Theta\left( cW_p - 1 \right) \right\rangle
\end{align}
where the step function $\Theta$ returns $1$ if $cW_p >1$ and $0$ otherwise. The tangling index therefore counts the number of obstacles that a worm winds around (Fig.~\ref{fig4}B), and is a measure of the mean degree of a tangle graph. Since connected graphs asymptotically have mean degree at least 2, we identify $\mathcal{T}^\ast \approx 2$ as the critical tangling index separating tangled states, with $\mathcal{T} > 2$, from loose states, with $\mathcal{T} < 2$. Near-critical trajectories (SI) bear a striking resemblance to curves solving the famous ``picture-hanging puzzle"~\cite{demaine2014picture}, which asks how to hang a picture on 2 pegs so that it falls if either peg is removed. Critical worm gaits could therefore be associated with such topological quick-release mechanisms. Indeed, our tomographic reconstructions indicate that worms form near-critical tangles (Fig.~\ref{fig2}F), thus balancing tangle stability with ability to disentangle rapidly.

The tangling index enables the topological state to be predicted from worm motion and spacing (Fig.~\ref{fig4}C). Assuming small noise terms (SI), the worm-head trajectories are characterized by speed $v$, turning rate $\alpha$ and angular switching rate $\lambda$, and $\ell$ captures the worm spacing. This leads to two dimensionless quantities, the chirality number $\gamma = \alpha/2\pi\lambda$ and the loop number, $R = v/\alpha \ell$, which measures the size of the loops produced by the worm trajectory in units of $\ell$. The resulting phase diagram, $\mathcal{T}(\gamma, R)$, explains the observed values of $\gamma$ and $R$ for worms executing tangling and untangling gaits (Fig.~\ref{fig4}C), and demonstrates that the loop number $R$ can also be used to control topological state. For example, larger values of $R$ allow a worm to wind around more obstacles, increasing topological complexity. However, for $R>0.5$, the chirality number $\gamma$ is the key determinant of tangle state (Fig.~\ref{fig4}C), indicating that tangle topology can be controlled purely by changing the rate $\lambda$ at which the turning direction switches. 3D simulations confirm the validity of this intuitive picture, demonstrating that by tuning $\gamma$, active filaments can be programmed to reversibly tangle and untangle, independently of head speed $v$. The phase diagram therefore reveals how tangle topology can be robustly controlled by manipulating only the chiral dynamics of the constituent filaments (Fig.~\ref{fig4}D,E; movie S3).

Blackworm locomotion lies close to the critical tangling threshold (Fig.~\ref{fig4}C), indicating that blackworm gaits are mechanically optimized for crossing the tangling-untangling barrier.
However, our mean-field tangling model predicts a large space of tangling and untangling strategies, within which blackworms occupy a relatively small region. Accounting for energetics helps identify the topological strategies which are inefficient for blackworms. For example, untangling with small $R$ requires forming small, energetically costly loops. Similarly, untangling using linear trajectories corresponding large $R$ gaits requires braids to be unravelled by pulling rather than unweaving, a motion which is associated with a higher friction penalty~\cite{sano2022exploring}. Furthermore blackworm dynamics are necessarily multifunctional, and topological requirements must be balanced with the need to support efficient, biologically feasible locomotion~\cite{drewes1989hindsight, kudrolli2019burrowing,wolgemuth2000twirling}. For example the helical waves of alternating chirality that promote untangling have also been identified in the context of worm swimming~\cite{drewes1993helical, drewes1999helical}. On the other hand, the highly entangled region of phase space with $\gamma>1, R>0.5$ suggests that there are stable tangle topologies not accessed by the worm tangles. Such a tangle would contain chiral filaments, in contrast to our observed living worm tangles (Fig.~\ref{fig1}G). The chirality number and loop number thus demonstrate how complex topologies may be created and tested beyond the biologically feasible regime.

\par
To conclude, active helical waves produced by the motion of individual worms facilitate collective tangling and ultrafast untangling. The fact that the underlying mechanisms are generic, and the predictions of elasticity theory are known to generalize across a wide range of scales~\cite{audoly2010elasticity}, raises the question of whether the results of our mean-field tangling model could apply to other systems of packed and tangled fibers~\cite{cantarella2020radius, grosberg1996flory, sulkowska2012conservation, chakrabarti2020trapping, weber2015random}. A particularly interesting class of systems involves tangled DNA~\cite{sato2021multistep,orlandini2019synergy,winkler2020physics,klotz2018motion}, where the key ingredients of activity and chirality, which underlie our model, are present. Mounting evidence that activity affects chromatin topology~\cite{saintillan2018extensile,banigan2019limits} underscores the importance of the kind of activity-topology interactions captured by our mean-field theory. More generally, the mechanism of fast untangling through head activity discovered here could provide design principles for a class of topological materials that can tangle and untangle on demand. Such tangle-based adaptive materials have recently found application in grasping problems~\cite{becker2022active}. Our model additionally demonstrates methods for fine control of tangle topology, opening up the possibility of programming a wide range of behaviors into a single topologically adaptive material by harnessing the large internal state space of tangles. The framework developed here could help understand the mechanical advantages of specific classes of tangles, and aid in the development of multifunctional materials based on topological properties.

\bibliography{references}

\textbf{Acknowledgements}
We thank Stanislav Emelianov (Georgia Tech) for sharing instrument facilities for ultrasound imaging. This work was supported by a MathWorks fellowship (V.P.P.),  a Stanford Science fellowship (V.P.P.), the NSF Graduate Research Fellowship Program (H.T.), a  Georgia Tech’s President’s Fellowship (H.T.), Georgia Tech’s Presidential Undergraduate Research Award (PURA) (E.K.) the MIT Mathematics Robert E. Collins Distinguished Scholar Fund (J.D.), Sloan Foundation Grant G-2021-16758 (J.D.). M.S.B. acknowledges funding support from NIH Grant R35GM142588; NSF Grants 1817334; 2218382; NSF CAREER 1941933; and the Open Philanthropy Project.

\end{document}


\title{Supplementary Materials:\\
Ultrafast reversible self-assembly of living tangled matter}
\author{Vishal P. Patil}
\thanks{V.P.P. and H.T. contributed equally to this work.}
\address{School of Humanities and Sciences, 
Stanford University,
450 Serra Mall, Stanford, CA 94305}

\author{Harry Tuazon}
\thanks{V.P.P. and H.T. contributed equally to this work.}
\address{School of Chemical and Biomolecular Engineering, Georgia Institute of Technology, Atlanta, GA 30318}

\author{Emily Kaufman}
\address{School of Chemical and Biomolecular Engineering, Georgia Institute of Technology, Atlanta, GA 30318}
\author{Tuhin Chakrabortty}
\address{School of Chemical and Biomolecular Engineering, Georgia Institute of Technology, Atlanta, GA 30318}

\author{David Qin}
\address{Wallace H. Coulter Department of Biomedical Engineering, Georgia Institute of Technology, Atlanta, GA 30332}

\author{J\"orn Dunkel}
\thanks{Correspondence to dunkel@mit.edu; saadb@chbe.gatech.edu}
\address{Department of Mathematics, 
Massachusetts Institute of Technology, 
77 Massachusetts Avenue, 
Cambridge, MA 02139}

\author{M. Saad Bhamla}
\thanks{Correspondence to dunkel@mit.edu; saadb@chbe.gatech.edu}
\address{School of Chemical and Biomolecular Engineering, Georgia Institute of Technology, Atlanta, GA 30318}

\date{\today}

\maketitle

\section*{Experimental methods}

\subsection{\textbf{Animals}}
We procured and reared blackworms as described in Tuazon, et al.~\cite{tuazon2022oxygenation}. We placed another culture of worms in a 10-gallon aquarium filled with pool filtered sand. Water is continuously filtered using an aquarium filter and cooled to 12$^{\circ}$C using a water chiller. 
Studies with blackworms do not require approval by an institutional animal care committee.  

\subsection{\textbf{Worm tangling and untangling experiments}}
We used a Leica MZ APO microscope (Heerbrugg, Switzerland) with an ImageSource DFK 33UX264 camera (Charlotte, NC) at 30FPS to record all worm tangling experiments placed in a 15mm confocal glass bottom petridish. To encourage full entanglement, we placed worms that were kept in the aquarium into room temperature water, which varied from 21.0±0.3$^{\circ}$C. A Chronos color (Burnaby, British Columbia, Canada) high speed camera captured all disentangling experiments at 1067FPS in a 35mm petridish. To stimulate rapid disentanglement, we applied a quick 1-2s shock using a 9V battery. This method does not cause harm to the worms. The worms were chosen from a population with mean length $30\,$mm (2.s.f) and mean radius $0.3\,$mm (1.s.f). Untangling data were collected from 5 trials each containing 5 worms. Every worm was tracked yielding 25 trajectories. Tangling data were collected from 4 trials each containing 5 worms. In total, 18 of the trajectories were tracked.

\subsection{\textbf{Worm tracking}}
Recordings were converted into AVI format using Adobe$^\copyright$ Premiere Pro and were then imported into ImageJ~\cite{schneider2012nih} for image analysis. After enhancing the contrast, we applied background subtraction and converted stacks to 8-bit. We manually tracked the prostomium of the worms using TrackMate~\cite{tinevez2017trackmate} for the entirety of each stack or when we can no longer see the anterior segments. This occurs when worms are entangled or escape outside the camera's field of view. 

\subsection*{Worm Ultrasound methods}
\begin{figure*}[t]
	\centering
	\includegraphics[width=\textwidth]{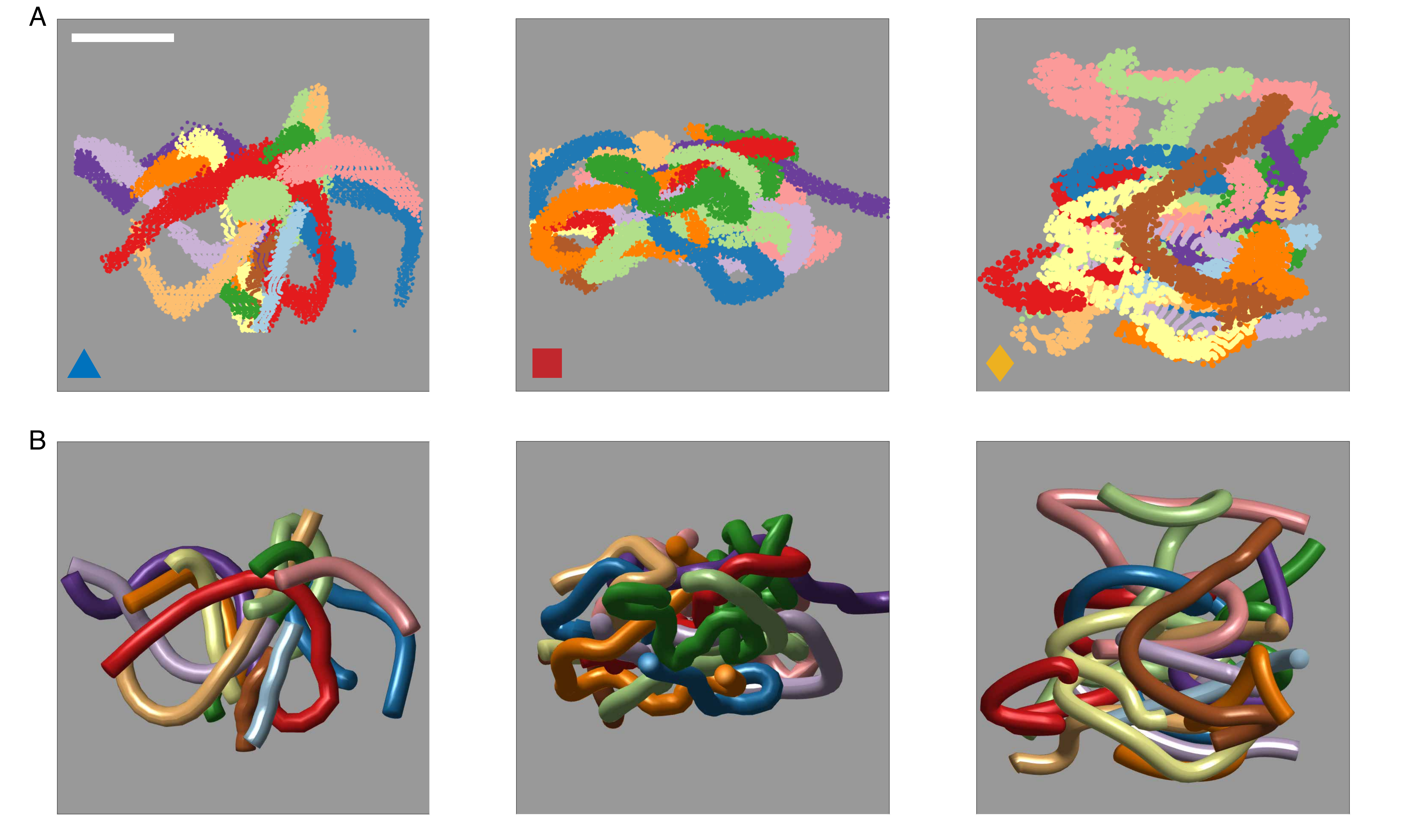}
	\caption{\textbf{Segmented and smoothed ultrasound data}
	(A)~Segmented ultrasound data for 3 different experiments with 12 worms each show tightly packed worms exhibiting thigmotactic behavior. Markers correspond to markers in Fig.~1H,I. Data set 3 (yellow diamond) is used as an initial condition for an untangling simulation (Fig.~3E and movie S2). Scale bar 5mm.
	(B)~Using the diffusion maps algorithm, curves are fit to the point clouds representing each worm in (A).}
	\label{SI_ultrasounds}
\end{figure*}

Worms were fixed in a 16$\%$ gelatin solution using filtered water and Type A gelatin (MP Biomedicals, Solon, OH). The solution was allowed to cool to 50$^{\circ}$C before being placed in a desiccator vacuum chamber for 5 minutes to remove any bubbles. The solution was then placed directly onto worms in a petridish and was allowed to solidify for at least 1 hour. After immobilization, the sample was transferred onto a flat gelatin disk; we then imaged them using a Vevo 2100 ultrasound system (VisualSonics, Toronto, Canada) and MS250 transducer (21 MHz center frequency), acquiring 2D grayscale ultrasound images at a 30Hz frame rate. The transducer was translated in the elevational direction to yield a 3D image of the worm tangle (Fig.~1B). Contrast was adjusted within the ultrasound system; the acquired data were then processed in MATLAB 2021b (Mathworks, Natick, MA). The 3D images underwent denoising and manual segmentation on MATLAB’s in-built GUI ‘Volume Segmenter’ to yield the final worm tangle representation (Fig.~1C). The worm centerline curves were extracted using the diffusion maps algorithm (Fig.~\ref{SI_ultrasounds}A,B).

\section*{Mathematical modelling}

\subsection*{Geometry and topology of curves}

In this section, we introduce the key geometrical and topological properties of curves which are discussed in the main text, including link and contact link.

\subsubsection*{Chirality}

To define chirality, we introduce the Frenet frame. Let $\gamma(s)$ be a curve parametrized by arc length with $\mathbf{t} = \gamma'$, so $|\mathbf{t}| = 1$. The Frenet frame is an orthonormal frame, $\{\mathbf{t}, \mathbf{n}, \mathbf{b}\}$, defined by
\begin{align*}
\bs{t}'(s) = \kappa\mathbf{n} \qquad \bs{n}'(s) = -\kappa\mathbf{t} + \tau\mathbf{b} \qquad \bs{b}'(s) = - \tau \mathbf{n}
\end{align*}
where primes denote $s$ derivatives. The quantities $\kappa(s)$ and $\tau(s)$ are the geometrical curvature and the torsion of the curve, respectively. $\kappa$ is related to the bending strain of an elastic fiber. In particular, if a cylindrical elastic fiber with radius $h$ has a centerline with curvature $\kappa(s)$, then the bending strain is $\kappa h$ (Fig.~1F).

We define the chirality, $\chi$, of the 3D curve by  
\begin{align*}
\chi = \bs{t} \cdot (\bs{t}' \times \bs{t}'' ) &= \bs{t} \cdot \left( \kappa \bs{n} \times (\kappa' \bs{n} + \kappa \bs{n}' ) \right)\\
&=  \bs{t} \cdot \left( \kappa\bs{n} \times (\kappa' \bs{n} -\kappa^2 \bs{t} + \kappa \tau \bs{b} ) \right)\\
&= \kappa^2 \tau
\end{align*}
This can be written in terms of the tangent vectors close to a given point
\begin{align*}
\bs{t}(s) \cdot \left(\bs{t}(s+h) \times \bs{t}(s - k) \right)  &= \bs{t} \cdot \left( (\bs{t} + h \bs{t}' + \frac{1}{2} h^2  \bs{t}'') \times (\bs{t} -k \bs{t}'  + \frac{1}{2} k^2  \bs{t}'') \right) \\
&= \bs{t} \cdot (\bs{t}' \times \bs{t}'' ) hk(h+k)/2
\end{align*}
In 2D, the signed curvature plays a similar role to chirality
\begin{align*}
    \chi_2 = \mathbf{t} \wedge \mathbf{t}' = \kappa 
\end{align*}

\subsubsection*{Link}

\begin{figure*}[t]
	\centering
	\includegraphics[width=\textwidth]{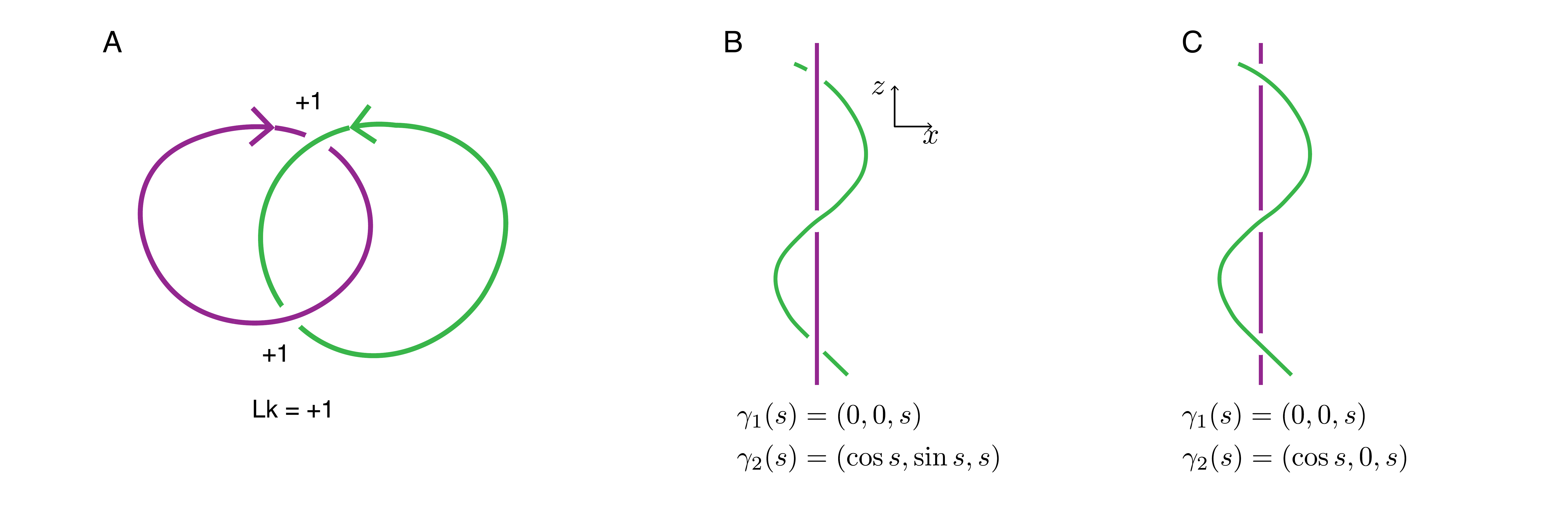}
	\caption{\textbf{Link and contact link}
	(A)~The link of 2 closed curves is given by the signed crossing number of a planar projection.
	(B,C)~The link of open curves measures topological obstruction. In (B), the green curve winds around the purple curve, and the resulting configuration has nonzero link. In (C), the green curve lies on top of the purple curve and the configuration has zero link.}
	\label{SI_topology}
\end{figure*}

Here we define the linking number of two closed curves both diagrammatically and in terms of the Gauss linking integral. We then show how these definitions can be extended to give linking numbers for open curves, as in Ref.~\cite{qinami2017quantifying}.

Link is a property of two closed curves. Consider two such closed curves $\gamma_1$, $\gamma_2$
\begin{align*}
\gamma_i: [0,L_i] \rightarrow \mathbb{R}^3
\end{align*}
The linking number can be defined in terms of the number of crossings, counted with sign, between $\gamma_1$ and $\gamma_2$ in a projection of the curves onto a plane (Fig.~\ref{SI_topology}A). Consider a projection, $P_{\bs{n}}$ in the direction $\bs{n}$
\begin{align*}
P_{\bs{n}} = I - \bs{n} \bs{n}^\top
\end{align*}
where $\bs{n}$ is a unit vector. The resulting planar diagram will have some number of crossings between the curves $\gamma_1$ and $\gamma_2$. Suppose there are $k$ crossings, at points $\gamma_1(s_i), \gamma_2(\sigma_i)$, where $i=1,2,...,k$. These crossing points satsify
\begin{align*}
\Big(\gamma_1(s_i) - \gamma_2(\sigma_i)\Big)  \parallel \bs{n}
\end{align*}
The sign of each crossing is
\begin{align*}
\text{sgn}(i) = \text{sgn} \Big(  \left[ \gamma_1(s_i) - \gamma_2(\sigma_i) \right] \cdot \left[ \gamma_1 ' (s_i) \times \gamma_2 ' (\sigma_i) \right] \Big)
\end{align*}
The total signed crossing number of the planar diagram obtained from the projection $P_{\bs{n}}$ is then
\begin{align*}
c_{\bs{n}} = \sum_i \text{sgn}(i)
\end{align*}
For closed curves, this signed crossing number is independent of the projection vector $\bs{n}$, and so is equal to twice the linking number
\begin{align*}
Lk(\gamma_1, \gamma_2) = \frac{1}{2}c_{\bs{n}}
\end{align*}

The linking number can also be written as an integral by considering the following smooth map
\begin{align*}
\Gamma : [0,L_1] \times [0,L_2] &\rightarrow S^2\\
(s,\sigma) &\mapsto \frac{ \gamma_1(s) - \gamma_2(\sigma) }{ |\gamma_1(s) - \gamma_2(\sigma)| } 
\end{align*}
Link is the degree of the map $\Gamma$
\begin{align*}
Lk(\gamma_1, \gamma_2) = \deg \Gamma &= \frac{1}{4\pi} \int ds\,d\sigma \, \Gamma \cdot \left( \Gamma_s \times \Gamma_\sigma \right)  \\ 
&= \frac{1}{4\pi} \int ds\,d\sigma\, \frac{ (\gamma_1(s) - \gamma_2(\sigma)) \cdot ( \gamma_1'(s) \times \gamma_2'(\sigma) ) }{ |\gamma_1(s) - \gamma_2(\sigma) |^3}
\end{align*}
Therefore, for closed curves, we have
\begin{align*}
Lk(\gamma_1, \gamma_2)
= \frac{1}{4\pi} \int ds\,d\sigma\, \frac{ (\gamma_1(s) - \gamma_2(\sigma)) \cdot ( \gamma_1'(s) \times \gamma_2'(\sigma) ) }{ |\gamma_1(s) - \gamma_2(\sigma) |^3} = \frac{1}{2}c_{\bs{n}}
\end{align*}
where $c_{\bs{n}}$ is the signed crossing number defined above.

For open curves, we can still define link by the above integral, however its interpretation in terms of signed crossing numbers is different. In particular, the signed crossing number for open curves depends on the choice of projection. As in Ref~\cite{qinami2017quantifying}, the corresponding result is now
\begin{align*}
Lk(\gamma_1, \gamma_2)
= \frac{1}{4\pi} \int ds\,d\sigma\, \frac{ (\gamma_1(s) - \gamma_2(\sigma)) \cdot ( \gamma_1'(s) \times \gamma_2'(\sigma) ) }{ |\gamma_1(s) - \gamma_2(\sigma) |^3} = \frac{1}{8\pi} \int d^2\bs{n}\, c_{\bs{n}}
\end{align*}
The linking integral therefore gives the average signed crossing number over all projections. This is a measure of the topological obstruction one curve imposes on the motion of the other (Fig.~\ref{SI_topology}B,C). For example, using cylindrical polar coordinates, the configuration of Fig.~\ref{SI_topology}B, in which one curve winds around another, has
\begin{align*}
(\gamma_1(s) - \gamma_2(\sigma)) \cdot ( \gamma_1'(s) \times \gamma_2'(\sigma) ) = -\mathbf{e}_r(\sigma) \cdot \left( \mathbf{e}_z \times \mathbf{e}_{\theta}(\sigma) \right) = 1
\end{align*}
so $Lk\neq 0$. In Fig.~\ref{SI_topology}C, the two curves can be easily separated from each other
\begin{align*}
(\gamma_1(s) - \gamma_2(\sigma)) \cdot ( \gamma_1'(s) \times \gamma_2'(\sigma) ) =  \cos \sigma \, \mathbf{e}_x \cdot \left( \mathbf{e}_z \times \sin\sigma \, \mathbf{e}_x \right) =0
\end{align*}
and thus $Lk=0$.

\subsubsection*{Contact link}

The contact link defined in the main text is a property of two curves $\gamma_1$ and $\gamma_2$, and a thickness parameter, $r$
\begin{align}\label{contactlink}
cLk(\gamma_i, \gamma_j \,; r) = \begin{cases}
|Lk(\gamma_i, \gamma_j)| \quad &\text{if } \min_{s,t} |\gamma_1(s) - \gamma_2(t)| < 2r \\
0 \quad &\text{otherwise}
\end{cases}
\end{align}
In other words, $cLk = |Lk|$ if tubes with radius $r$, centered around $\gamma_1$ and $\gamma_2$ are in contact, and $cLk=0$ otherwise. For worms, the value of $r$ is naturally given by the worm radius, so for notational convenience, we often drop the explicit $r$-dependence of $cLk$. However the change in contact link with $r$ is reminiscent of persistence homology and so can also be a useful tangle quantity (Fig.~2F). 

Given $N$ tubes with fixed $r$, and centerline curves, $\gamma_1, \gamma_2,...\gamma_N$, we can construct the contact link matrix, $C_{ij}$
\begin{align*}
C_{ij} = cLk(\gamma_i, \gamma_j)
\end{align*}
The adjacency matrix $A_{ij}$ of the tangle graphs constructed in the main text follows from thresholding the contact link matrix
\begin{align*}
A_{ij} = \begin{cases}
1 \quad \text{if } C_{ij} \geq 1/2 \\
0 \quad \text{if } C_{ij} < 1/2
\end{cases}
\end{align*}
We can obtain measures of the complexity of the tangle from the matrices $A_{ij}$ and $C_{ij}$. In particular, we define the total contact link per worm $T_{c}$, and the mean degree $\langle d\rangle$ as follows
\begin{align}
T_{c}(\gamma_1, ...,\gamma_N) &= \frac{1}{N} \sum_{i,j} C_{ij} = \left\langle \sum_{j} C_{ij} \right\rangle \label{total_cLk_per_worm} \\
\langle d\rangle &= \frac{1}{N}\sum_{i,j} A_{ij} \label{mean_degree}
\end{align}
$T_c$ is therefore a measure of the average amount of tangling between one worm and the rest of the tangle. The mean degree can be thought of as the thresholded analogue of the total contact link per worm. The radius dependence of $T_c$ is shown in Fig.~2F. Since $T_c$ depends on the sharp contact threshold in \eqref{contactlink}, we smooth our plots of $T_c$ against time (Fig.~3F and Fig.~4E). 

For larger values of $T_c$ and $\langle d\rangle$, the tangle is more complex, and the tangle graph has more edges. When $T_c$ and $\langle d\rangle$ are small, the $N$ curves do not form a coherent tangle. and the tangle graph will be sparse and disconnected. We can therefore use $T_c$ and $\langle d \rangle$ to map tangling to a percolation problem. Since a connected graph on $n$ vertices has mean degree at least $2-1/n$, we expect $T_c>2$ to correspond to tangled states, and $T_c <2$ to correspond to untangled states, in the large $n$ limit. From our ultrasound reconstructions, we find (Fig.~2E,F)
\begin{subequations}\label{cLk_per_worm_data}
\begin{align}
    T_c &= 1.8, \; 2.1,\; 1.7\\
    \langle d\rangle &= 1.2, \; 1.7, \; 1.2
\end{align}
\end{subequations}
where the values correspond to the data sets in Fig.~\ref{SI_ultrasounds}, from left to right. We note that the low values of $\langle d \rangle$ are due to the fact that our ultrasound worm tangles do not have connected tangle graphs (Fig.~2D,E).

\subsection*{Elastic model}

We model each worm using the Kirchhoff equations~\cite{bergou2008discrete} together with internal damping and a hydrodynamic friction~\cite{patil2020topological, patil2020discharging}. The worm centerline is given by $\bs{x}(s,t)$ where $s\in [0,L]$ is an arc length parameter for the unstretched fiber and $L$ is the worm length. The worm has material frame $[\mathbf{d}_1, \mathbf{d}_2,\mathbf{d}_3]$ where $\bs{x}' \parallel \mathbf{d}_3$ and primes denote $s$-derivtatives. The twist density is given by $\theta' = \mathbf{d}_1' \cdot \mathbf{d}_2$. Similarly, we define the angular velocity component $\omega_3 = \dot{\mathbf{d}}_1 \cdot \mathbf{d}_2 $, which is related to $\theta'$ by the identity 
\begin{align*}
\omega_3' + \mathbf{d}_3 \cdot \left( \mathbf{d}_3' \times \dot{\mathbf{d}}_3 \right) = \dot{\theta}'
\end{align*}
The Kirchhoff equations, augmented by terms for damping, friction, contact and activity, are
\begin{subequations}\label{Kirchhoff}
\begin{align}
\bs{f}^{\text{elast}} + \bs{f}^{\text{con}}  + \bs{f}^{\text{fric}} + \bs{f}^{\text{act}}&=  \rho A \dot{\boldsymbol{v}} + \gamma A\boldsymbol{v} - \eta A \boldsymbol{v}'' \\
\tau^{\text{elast}} + \tau^{\text{fric}} &= \rho I_3 \dot{\omega}_3 + \gamma I_3\omega_3 - \eta I_3(\omega_3 \mathbf{t})''\cdot \mathbf{t}
\end{align}
\end{subequations}
where $\bs{v} = \dot{\bs{x}}$, $\rho$ is the density, $A$ is the cross sectional area, $I_3$ is the perpendicular cross sectional moment of inertia and $\gamma, \eta$ are damping coefficients, with dimensions
\begin{align*}
[\gamma] = ML^{-3}T^{-1} , \qquad\quad  
[\eta]  = ML^{-1}T^{-1}
\end{align*}
We assume the fiber has circular cross section with radius $h$, so $A = \pi h^2$ and $I_3 = \pi h^4 /2$. Here we focus on the case where the activity is localized at the head of the worm, so $\bs{f}^{\text{act}}=0$, but there is an active force at the boundary, $\mathbf{F}^{\text{act}}(L,t)$. To write the boundary conditions, let $\mathbf{F}^{\text{tot}}, T^{\text{tot}} $ be the total force and torque on the rod, including the viscous dissipation terms
\begin{align*}
\left(\mathbf{F}^{\text{tot}}\right)'  &= \bs{f}^{\text{elast}} + \bs{f}^{\text{con}}  + \bs{f}^{\text{fric}} + \eta A \bs{v}'' \\
\left(T^{\text{tot}}\right)'  &= \tau^{\text{elast}} + \tau^{\text{fric}} + \eta I_3(\omega_3 \mathbf{t})''\cdot \mathbf{t}
\end{align*}
Here $\mathbf{F}^{\text{tot}}, T^{\text{tot}} $ have units of force and torque, whereas the terms $\bs{f}^{\text{elast}}, \tau^{\text{elast}}$, etc. have units of force density and torque density. Then the boundary conditions describing a worm driven by an active force at the head are
\begin{align*}
\mathbf{F}^{\text{tot}}(0,t) &=  0 ,
\qquad\quad \mathbf{F}^{\text{tot}}(L,t) = \mathbf{F}^{\text{act}}(t) \\
T^{\text{tot}}(0,t) &= 0 ,
\qquad\quad T^{\text{tot}}(L,t) = 0 
\end{align*}
The forces in torques in equation \eqref{Kirchhoff} consist of an elastic force and torque arising from the elastic energy of the fiber, a contact force, $\bs{f}^{\text{con}}$, which prevents fibers from intersecting, a frictional force $\bs{f}^{\text{fric}}$ and torque $\tau^{\text{fric}} $. Following Ref.~\cite{bergou2008discrete}, our simulation framework stores a discretized $\bs{x}$ and $\theta$, and the elastic forces and torques are calculated by taking derivatives of a discretized elastic energy. The discretized contact and friction forces are described further in Ref.~\cite{patil2020topological} and follow continuum expressions described below.

\bigskip

We can describe the elastic forces and torques on the left hand side of equation \eqref{Kirchhoff} in more detail. The elastic forces and torques are derivatives of the the elastic energy of the fiber, which is made up of the bending energy $\mathcal{E}_b$, twisting energy, $\mathcal{E}_{\text{tw}}$ and stretching energy $\mathcal{E}_s$
\begin{align*}
\mathcal{E}_b = \frac{1}{2} E_b I  \int_0^L ds \, \kappa^2 , 
\qquad\quad  \mathcal{E}_{tw} = \frac{1}{2} \mu_b J \int_0^L ds\, \theta'^2 , 
\qquad\quad \mathcal{E}_s = \frac{1}{2} EA \int_0^L ds\, \left( |\bs{x}'| - 1 \right)^2
\end{align*}
where $I$ is the cross sectional moment of inertia, $J$ is the moment of twist, $E$ is the Young's modulus of the fiber, $E_b$ is the bending modulus, and $\mu_b$ is the shear modulus. Under the assumption of circular cross sections, $I = \pi h^4/4$ and $J = \pi h^4/2$. Owing to the complex internal structure of worms, we take $E_b \neq E$. The Poisson's ratio, $\nu$, relates the shear modulus to the bending modulus, $\mu_b = E_b/(2+2\nu)$. 
The elastic forces and torques are then
\begin{align*}
\bs{f}^{\text{elast}} (s) &= - \frac{\delta \mathcal{E}_b}{\delta \bs{x}(s)} - \frac{\delta \mathcal{E}_{tw}}{\delta \bs{x}(s)} - \frac{\delta \mathcal{E}_s}{\delta \bs{x}(s)} \\
\tau^{\text{elast}}(s) &= - \frac{\delta \mathcal{E}_{tw}}{\delta \theta(s)}
\end{align*}

The remaining forces describe self-interactions within the fiber. There are many different approaches to discretizing and implementing these interactions~\cite{patil2020topological,tong2022fully,bergou2008discrete, bergou2010discrete}. We begin with the contact forces, which prevent the fiber from intersecting itself. Let $\bs{r}(s,\sigma) = \bs{x}(s) - \bs{x}(\sigma)$ and $\hat{\bs{r}} = \bs{r} / | \bs{r} |$. To quantify contact, set
\begin{align*}
p(s,\sigma) = 1 - \frac{\bs{r}(s,\sigma)}{2h_{\text{eff}}}
\end{align*}
where $h_{\text{eff}} - h$ is small. If $h_{\text{eff}}=h$, then $p=0$ when $\bs{x}(s), \bs{x}(\sigma)$ are touching and $p=1$ when $\bs{x}(s), \bs{x}(\sigma)$ overlap completely. Choosing $h_{\text{eff}} \neq h$ has the desirable effect of smoothing out the contact region and the associated contact forces~\cite{tong2022fully}. When $h_{\text{eff}}>h$, we use the threshold $h_{\text{eff}}$ to calculate the the contact link \eqref{contactlink} from simulation data. Using $p$ we can construct a contact potential energy, $V(p)$
\begin{align*}
V(p) &= \frac{K}{2} p^2 + K p_0 \left( \frac{1}{4} p^4 + \frac{1}{6} p^6 + \frac{1}{8} p^8 \right) \quad \text{for } p>0 \\
V(p) &= 0 \quad \text{for } p<0
\end{align*}
where $K$ is an effective bulk modulus, and $p_0$ is a constant that sets the rate at which the potential stiffens. Here, we take $p_0 = 100$. The contact force is then
\begin{align*}
\bs{f}^{\text{con}}(s) = -\int_{|\sigma - s| > 2h_{\text{eff}}} d\sigma \, \bs{x}'(s) \cdot \bs{x}'(\sigma)\, \frac{dV}{dp} \bigg|_{p(s,\sigma)} \,  \hat{\bs{r}}(s, \sigma)
\end{align*}
The integral excludes points $\bs{x}(\sigma)$ which are close to $\bs{x}(s)$ along the fiber, i.e. points with $|\sigma - s| \leq 2h_{\text{eff}}$. This ensures that only points $\bs{x}(s), \bs{x}(\sigma)$ which are legitimately in contact are included in the contact force integral.

We implement the contact law by discretizing the rod~\cite{bergou2008discrete,patil2020topological,tong2022fully} into vertices, $\bs{x}_i = \bs{x}(s_i)$, connected by links with midpoints given by $\bar{\bs{x}}_i = (\bs{x}_i + \bs{x}_{i+1})/2$. An appropriate choice of $h_{\text{eff}}$ depends on the discretization length of the rod, $|\bar{\bs{x}}_i |$. Contact forces as described above are then implemented for link-link interactions and vertex-vertex interactions. The length scale $h_{\text{eff}}$ is chosen for each interaction type so that the average value of $h_{\text{eff}}$ is $h$. In our $30$-worm, $19$-worm (Fig.~4D, movie S3) and $8$-worm (Fig.~3E movie~S2) simulations, we have $|\bar{\bs{x}}_i | = 2h$ and we take $h_{\text{eff}} = 1.25 h$ for link-link interactions, and $h_{\text{eff}}=0.75$ for vertex-vertex interactions.

Finally, we include hydrodynamic-type friction terms~\cite{patil2020topological}
\begin{align*}
\bs{f}^{\text{fric}}(s) &= -\zeta A \int_{|\sigma - s| > 2h}  d\sigma \, \Theta\left[ p(s,\sigma) \right] \frac{ \dot{\bs{x}}(s) - \dot{\bs{x}}(\sigma) } { |\bs{r}(s, \sigma) |^c } \\
\tau^{\text{fric}}(s) &=   \zeta_{tw} J \int_{|\sigma - s| > 2h}  d\sigma \, \Theta\left[ p(s,\sigma) \right]  \frac{ \bs{r}(s,\sigma) \times \left( \dot{\bs{x}}(s) - \dot{\bs{x}}(\sigma) \right) \cdot \mathbf{d}_3(\sigma) }{  |\bs{r}(s, \sigma) |^{c_{tw}}}
\end{align*}
Here, following Ref.~\cite{patil2020topological} we take $c=2$ and $c_{tw} = 4$, so the coefficients $\zeta, \zeta_{tw}$ have the same dimensions
\begin{align*}
[\zeta] = [\zeta_{tw}] = M L^{-2} T^{-1}
\end{align*}
We set $\zeta=\zeta_{tw}$.

\subsection*{Worm head trajectories}

In this section, we construct a stochastic model for worm head trajectories in 2D, which we then use to formulate a stochastic, 3D active force, $\mathbf{F}^{\text{act}}(t)$, that drives worms. Coupling this active force to the elastic model described above \eqref{Kirchhoff} can then be shown to produce tangling and untangling dynamics (movies S2 and S3).

\subsubsection*{2D Worm head trajectories}

Our 2D stochastic model for the worm head trajectories is based on experimental data (Fig.~3). We assume the worm head has average speed $v$, and turns at average angular speed $\alpha$. The turning direction switches from clockwise to anticlockwise at rate $\lambda$. These speeds are augmented by stochastic terms governed by a translation diffusivity $D_T$ and rotational diffusivity $D_R$
\begin{subequations}\label{SDE2}
\begin{align}
d\mathbf{X} &= v \mathbf{N}_\theta \, dt + \sqrt{2D_T}  * d\mathbf{B}(t)\\
d\mathbf{\theta} &= (-1)^{S(t)} \alpha \, dt + \sqrt{2D_R} * d W(t) 
\end{align}
\end{subequations}
where $\mathbf{B}(t)$ and $W(t)$ are independent Brownian processes, and $S(t) = \hat{S}(t) + S_0$, where $\hat{S}(t)$ is a Poisson process of rate $\lambda$, i.e. $\hat{S}(t) \sim Po(\lambda t)$, and $S_0 \sim Ber(1/2)$ is independent of $\hat{S}(t)$. The process $S(t)$ is then effectively a rate $\lambda$ Poisson process with random initialization, so $S(t)-S(0) \sim Po(\lambda t)$. In particular,
\begin{align*}
\mathbb{E}\left[ (-1)^{S(t)} \right] = \mathbb{E}\left[ (-1)^{\hat{S}(t)+S_0}  \right] = \mathbb{E}\left[ (-1)^{\hat{S}(t)} \right] \mathbb{E}\Big[ (-1)^{S_0} \Big] = 0
\end{align*}
In Langevin notation, writing $\bs{x} = \mathbf{X}$ and $\bs{n}_{\theta} = \mathbf{N}_{\theta}$, equation \eqref{SDE2} becomes
\begin{subequations}\label{langevin}
\begin{align}
\dot{\bs{x}} &= v\bs{n}_{\theta} + \bs{\xi}_T(t)\\
\dot{\theta} &= (-1)^{S(t)}\alpha +  \xi_R(t)
\end{align}
\end{subequations}
where $\xi_T, \xi_R$ are independent white noise processes, satisfying 
\begin{align*}
\langle \left(\xi_T\right)_i(t) \left(\xi_T\right)_j(t') \rangle = 2D_T \delta_{ij} \delta(t-t') , \qquad \langle \xi_R(t) \xi_R(t') \rangle = 2D_R \delta(t-t')
\end{align*}
where angle brackets denote expectations. Assuming the noise terms are small, the trajectory shapes are determined by $\alpha, \lambda$ and $v$. In particular, the chirality number, $\gamma = \alpha/2\pi\lambda$ sets the number of full loops the head makes before switching direction, and $v/\alpha$ sets the size of these loops. Although the noise terms in equation \eqref{SDE2} are taken to be Brownian, the nature of the noise in the real, biological system is expected to be more complex. However, the choice of Brownian noise makes this model more analytically tractable.

\subsubsection*{Rotation index of 2D trajectories}

When $\theta(0) = 0$, the value of $\theta$ is equal to the total winding of the velocity vector around the origin, and is known as the rotation index, $I_R$, of the curve
\begin{align*}
    I_R = \frac{1}{2\pi}\theta(t)
\end{align*}
Although this is distinct from the winding number of a trajectory about a point, intuitively, we expect that trajectories with larger values of $I_R$ should tangle more readily. We can find the $I_R$ by calculating moments of $\theta$. Under the assumption of Brownian noise.
\begin{align*}
\mathbb{E}[\theta(t)] = \alpha \int_0^t dt'\, \mathbb{E}\left[(-1)^{S(t')}\right] = 0
\end{align*}
Using Ito's lemma
\begin{align*}
d(\theta^2) = 2 \left((-1)^{S(t)} \alpha \theta + D_R\right) \,dt  +2\theta\sqrt{2D_R} *dW(t)
\end{align*}
Thus
\begin{align*}
\frac{d}{dt} \mathbb{E}[\theta^2] &= 2D_R+ 2\alpha \mathbb{E} [ (-1)^{S(t)} \theta ] \\
&= 2D_R + 2\alpha^2 \mathbb{E} \left[ (-1)^{S(t)} \int_0^t dt' \, (-1)^{S(t')} \right] \\
&= 2D_R + 2\alpha^2 \int_0^t dt' \, \mathbb{E} \left[ (-1)^{S(t)-S(t')+2S(t')} \right] \\
&= 2D_R + 2\alpha^2 \int_0^t dt' \, e^{-2\lambda (t-t')} \\
&= 2D_R + 2\alpha^2 e^{-2\lambda t} \frac{1}{2\lambda} \left( e^{2\lambda t} -1 \right)\\
&= 2D_R + \frac{\alpha^2}{\lambda}\left( 1 - e^{-2\lambda t} \right)
\end{align*}
Integrating this gives
\begin{align*}
\mathbb{E}\left[\theta^2\right] = \left(2D_R + \frac{\alpha^2}{\lambda} \right) t - \frac{\alpha^2}{2\lambda^2} \left( 1 - e^{-2\lambda t}\right)
\end{align*}
We can simplify this expression by using the chirality number, $\gamma = \alpha/2\pi \lambda$, which counts the typical number of loops of a given handedness formed by the trajectory before it switches turning direction. In particular, the rotation index satisfies
\begin{align*}
    \mathbb{E}\left[I_R^2\right] = \frac{1}{4\pi^2} \mathbb{E}\left[ \theta^2 \right]  = \frac{1}{2\pi}\left(\frac{D_R}{\pi\alpha} + \gamma \right) \alpha t - \frac{1}{2}\gamma^2 \left( 1 - e^{-\alpha t / \pi \gamma}\right)
\end{align*}
Defining the dimensionless time $\tilde{t} = \alpha t$, and the we can simplify this expression further
\begin{align}\label{rotation_index}
    \mathbb{E}[I_R^2] = \frac{1}{2\pi}\left(\frac{D_R}{\pi \alpha} + \gamma \right) \tilde{t} - \frac{1}{2}\gamma^2 \left( 1 - e^{-\tilde{t} / \pi \gamma}\right)
\end{align}
This is an increasing function of $\gamma$, suggesting that trajectories with larger $\gamma$ have a greater propensity to tangle.

\subsubsection*{3D Worm head dynamics}

3D head dynamics can be built from this 2D model by adding a perturbation in the $z$-direction. To mimic the trajectories described above, we introduce an active force $\mathbf{F}^{\text{act}}$ at the head of each worm which rotates at rate $\alpha$ and switches rotation direction at rate $\lambda$. More concretely, setting $\mathbf{F}^{\text{act}} = q \bs{n}_{\theta}$ where $\bs{n}_{\theta} = (\cos\theta, \sin\theta,0)$, we define the following dynamics for the active head force
\begin{subequations}\label{active_force}
\begin{align}
\dot{\theta}(t) &=  (-1)^{S(t)}\alpha + \xi_R(t) \\
\mathbf{F}^{\text{act}}(t) &=  q \bs{n}_{\theta}(t) + \bs{\xi}_F(t)
\end{align}
\end{subequations}
The $\theta$ equation and the Poisson process $S(t)$ are as in \eqref{langevin}, and $\bs{\xi}_F$ in the force equation is a white noise process in 3D satisfying
\begin{align*}
\langle \left(\xi_T\right)_i(t) \left(\xi_T\right)_j(t') \rangle = 2D_F \delta_{ij} \delta(t-t') 
\end{align*}
where $D_F$ has units $[D_F] = M^2 L^2 T^{-3}$. The noise term therefore acts as a perturbation, moving the filaments out of plane. In practice, a combination of the noise term and contact effects produce the 3D tangled structures we see in simulations (movies S2 and S3). The parameters $\alpha$ and $\lambda$ are explicit in these active force equations. The speed $v$ follows from considering the equations of motion for the elastic fiber, and can be solved for numerically using our simulation framework.

\subsection*{Mean-field tangling model}

Using trajectory equations for the worm head, we can compute how much a single worm tangles around a fixed set of obstacles, which represent the structure of a fixed background tangle. This idea leads to a mean-field theory of tangling. Although our numerical tests of this theory will necessarily use the 3D dynamics described above (equation \ref{active_force}), for simplicity, we will build the mean-field model in 2D, using the SDEs described in equation \eqref{SDE2} and \eqref{langevin}
\begin{align*}
\dot{\bs{x}} &= v\bs{n}_{\theta} + \bs{\xi}_T(t)\\
\dot{\theta} &= (-1)^{S(t)}\alpha +  \xi_R(t)
\end{align*}
We further assume that $D_T$ is negligible. As a proxy for tangling, we can construct a tangling parameter based on the winding of the curve $\bs{x}(t)$ around a specified set of points $\Lambda \subset \mathbb{R}^2$. The set $\Lambda$ introduces another length scale into our model. We let $\ell$ be the characteristic spacing between points in $\Lambda$. The winding number of $\bs{x}(t)$ about $p\in \Lambda$ in the time taken to travel one worm length, $t = L/v$ is
\begin{align*}
W_p(\bs{x}) &= \frac{1}{2\pi} \int_0^{L/v} dt\; \frac{d}{dt} \arg \left( \bs{x}(t) - p\right) \nonumber
\end{align*}
Since tangling depends on geometrical contact as well as topological winding, we keep track of both the winding and contact around obstacles $p\in\Lambda$. We define the contact winding, $cW_p(\bs{x})$, in the same way as contact link \eqref{contactlink} 
\begin{align}\label{contact}
cW_p(\bs{x}) = \begin{cases}
|W_p(\bs{x})| \quad &\text{if } \min_{t \in [0, L/v]} |\bs{x}(t) - p| < r \\
0 &\text{otherwise}
\end{cases}
\end{align}
where $r$ is the contact threshhold, so the curve is said to be in contact with a point $p \in \mathbb{R}^2$ if it comes within distance $r$ of $p$. In addition, we threshhold the winding number, so that winding around a point $p\in\Lambda$ is only counted if $cW_p\geq 1$. This is necessary to exclude certain configurations where $W_p$ is non-negligible but the interaction between $\bs{x}$ and $p$ is marginal. For example, for $\bs{x}(t) = (\epsilon, vt - L/2)$, and $p=0$, we have
\begin{align*}
W_p(\bs{x}) = \frac{1}{\pi}\arctan\left(\frac{L}{2\epsilon}\right)
\end{align*}
Thus $W_p$ can approach $1/2$ even though the curve $\bs{x}$ lies adjacent to the point $p$. This motivates the following tangling parameter, which combines the topological information of winding with the geometric information of contact
\begin{align}\label{tangling_index}
\mathcal{T} = \left\langle \sum_{p \in \Lambda} \Theta\left( cW_p -1 \right) \right\rangle
\end{align}
where $\Theta$ is the Heaviside step function. The obstacles in $\Lambda$ can be thought of as other worms in the tangle. The tangling index therefore counts the number of worms a given worm interacts with topologically. In this picture, the tangling index can be viewed as the mean-field analogue of the contact link based tangle quantifiers constructed earlier. In particular, $\mathcal{T}$ is constructed similarly to the total contact link per worm (equation \ref{total_cLk_per_worm}) and the mean degree of the tangle graph (equation \ref{mean_degree}). This motivates a connection with percolation theory, as discussed above. We assume that a tangled state has a connected tangle graph and ask which values of $\mathcal{T}$ correspond to dynamics (equation \ref{active_force}) capable of producing tangled states. Since a connected graph on $n$ vertices has mean degree at least $2-2/n$, we conjecture that $\mathcal{T}^\ast \approx 2$ is the critical value separating tangled states from untangled states. This agrees with our experimental data (Fig.~4C). In addition, our tomographic reconstructions have total contact link per worm approximately equal to 2 (Fig.~2F, and equation \ref{cLk_per_worm_data}), suggesting that living worm tangles are tuned for ease of untangling.

\subsubsection{Analysis of tangling index}

\begin{figure*}[t]
	\centering
	\includegraphics[width=\textwidth]{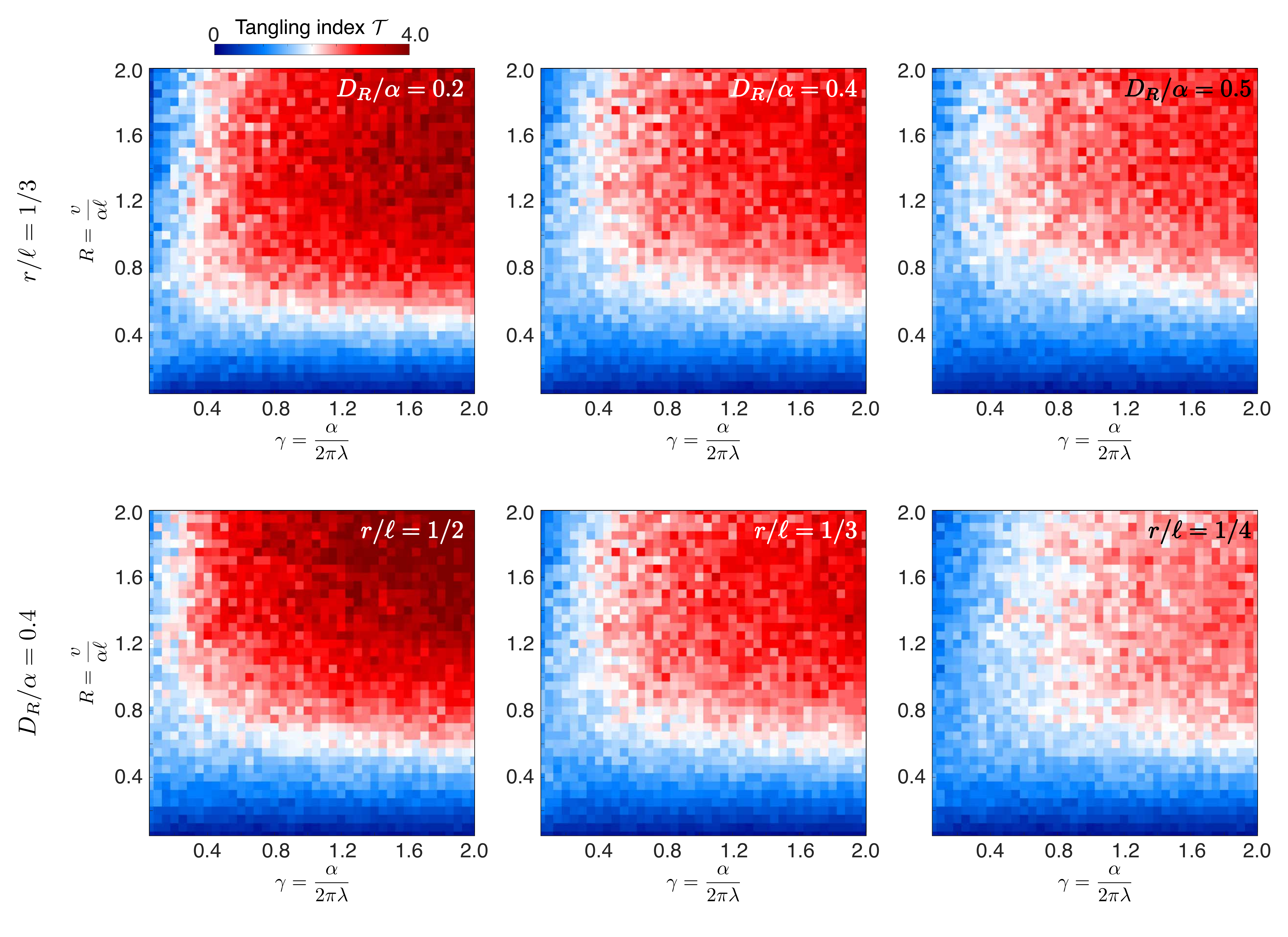}
	\caption{\textbf{Effect of noise and contact threshold on mean-field phase diagram}
    The phase diagrams depend weakly on noise and contact threshold. In Fig.~4C, we choose $r/\ell = 1/3$ and $D_R/\alpha = 0.4$ (middle column).  }
	\label{SI_phase_diagrams}
\end{figure*}

We can write $\mathcal{T}$ as a function of dimensionless parameters by non-dimensionalizing the the trajectory equations \eqref{langevin}. Let $\tilde{t} = \alpha t$ and $\tilde{\bs{x}} = \bs{x} / \ell$, where $\ell$ is the characteristic spacing between points in $\Lambda$. Then equation $\eqref{langevin}$ becomes
\begin{subequations}\label{SDEnondim}
\begin{align}
\frac{d\tilde{\bs{x}}}{d\tilde{t}} &= R \bs{n}_{\theta}  \\
\frac{d\theta}{d\tilde{t}} &= (-1)^{\tilde{S}(\tilde{t})}  + \tilde{\xi}_R(\tilde{t})
\end{align}
\end{subequations}
where $R = v/\alpha\ell$ is the loop number, $\tilde{S}(\tilde{t}) = S(\tilde{t}/\alpha)$, the rotational noise term $\tilde{\xi}_R(\tilde{t}) = \alpha^{-1}\xi_R(\tilde{t}/\alpha)$ satisfies
\begin{align*}
    \left\langle \tilde{\xi}_R(\tilde{t})\tilde{\xi}_R(\tilde{t}') \right\rangle  = \frac{2D_R}{\alpha}\delta(\tilde{t}-\tilde{t}')
\end{align*}
and we have assumed that the translational noise coefficient, $D_T$, is negligible. The process $\tilde{S}(\tilde{t})$ is a Poisson process with dimensionless rate $\lambda/\alpha = 1/2\pi\gamma$ where $\gamma$ is the chirality number
\begin{align*}
\tilde{S}(\tilde{t}) - \tilde{S}(0) \sim Po \left( \frac{\tilde{t}}{2\pi\gamma} \right)
\end{align*}
We therefore find three dimensionless parameters from $\eqref{SDEnondim}$, the chirality number $\gamma$, the loop number $R$ and a non-dimensional noise, $D_R/\alpha$. To calculate $\mathcal{T}$, we run the process for time $t = L/v$ and quantify contact using the contact threshold $r$ (equation \ref{contact}). The gives two additional dimensionless parameters, $L/\ell$ and $r/\ell$. For a given obstacle set, $\Lambda$, the tangling index can therefore be written as
\begin{align}
    \mathcal{T} = \mathcal{T} \left( \gamma, R \,; \frac{D_R}{\alpha}, \frac{r}{\ell} , \frac{L}{\ell} \right)
\end{align}
In this study, we have focused on the role of the dynamical dimensionless parameters, $\gamma$ and $R$, which arise from equation \eqref{SDE2}. The effect of the noise and contact parameters can be explored numerically (Fig.~\ref{SI_phase_diagrams}) and do not change the underlying structure of the phase diagram. In Fig.~4C, we choose $D_R/\alpha = 0.4$ and $r/\ell = 1/3$. The remaining degrees of freedom, $L/\ell$ and $\Lambda$ govern the placement and spacing of winding obstacles.

The worm length, $L$, affects $\mathcal{T}$ by setting the time for which the process $\mathbf{X}$ is run. We set $L/\ell = 25$ in the phase diagrams shown here (Figs.~4C, \ref{SI_phase_diagrams}-\ref{SI_cutoff_frequencies}). Larger $L$ will increase $\mathcal{T}$, since the curve $\mathbf{X}$ will have more time to wind around points. This matches with our intuition that longer filaments tangle more easily. From ultrasound data, we find $\ell = 3h$, and $L/h \approx 80$, where $h$ is worm radius, which gives $L/\ell\approx 27$. Note that this implies that the chosen contact threshold, $r$, is less than $2h$. This can be understood as a consequence of the fact that our mean-field model is treating a dynamic 3D tangle as a static 2D point cloud. For example, a worm of radius $h$ cannot pass between two parallel worms separated by $\ell=3h$, without causing some deformation. The fact that $r<2h$ is due to the fact that our model does not explicitly account for such deformation effects. In particular, our mean-field model does not explicitly depend upon the worm radius $h$, although worm radius effects are captured by the contact threshold $r$.

The set of obstacles, $\Lambda$ acts as the background tangle through which a worm moves. Although this background could be amorphous, for simplicity, here we assume that $\Lambda$ has a lattice structure. Numerical investigations further reveal that the tangling index $\mathcal{T}$, does not strongly depend on lattice type (Fig.~\ref{SI_mean_field_curves}A,B). This validates our assumption that $\Lambda$ affects $\mathcal{T}$ mostly through the lattice spacing $\ell$.

\subsubsection*{Mean-field trajectories through $\Lambda$}

\begin{figure*}[t]
	\centering
	\includegraphics[width=\textwidth]{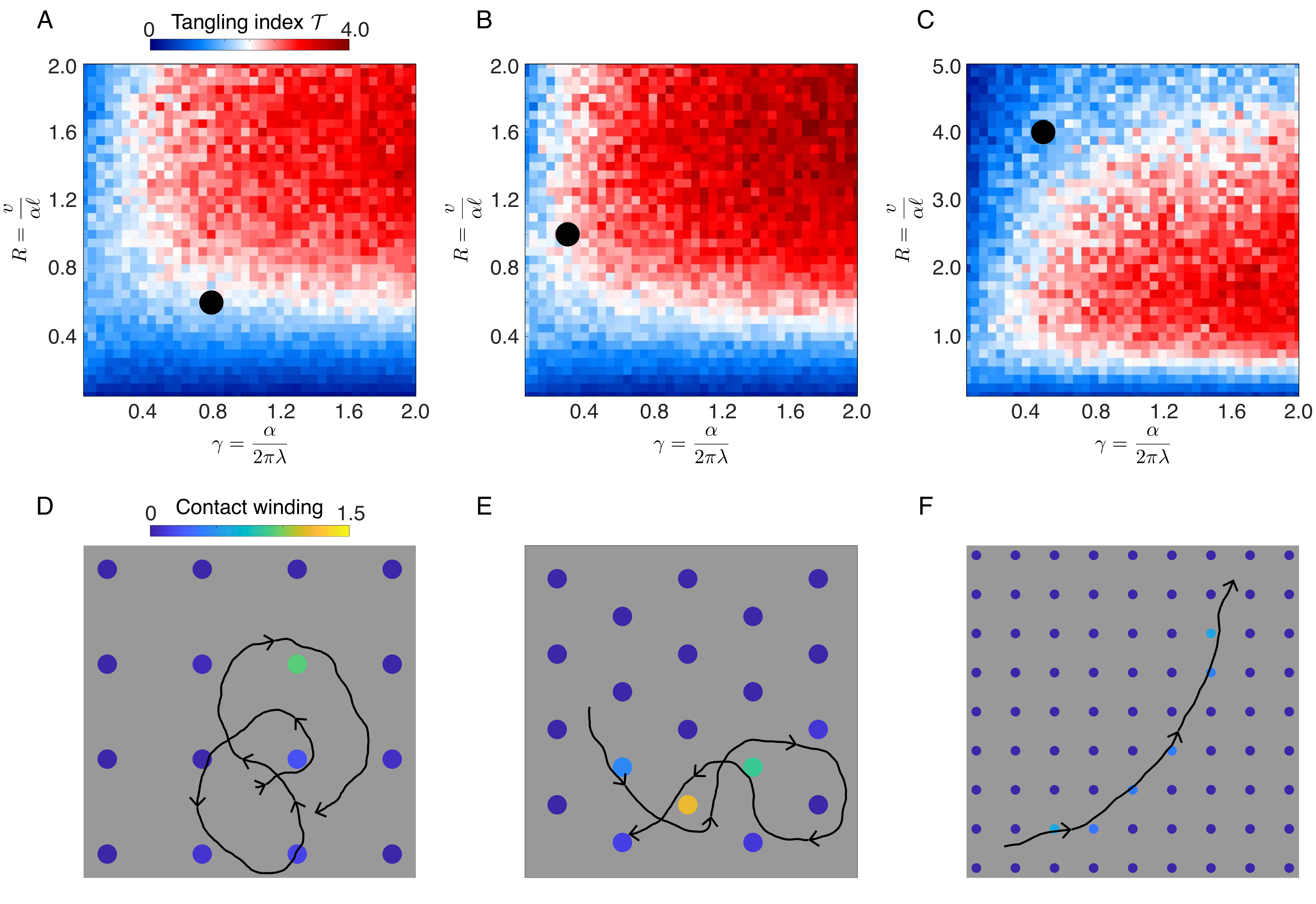}
	\caption{\textbf{Effect of obstacle placement and spacing.}
    (A,B)~Choosing a square (A) or triangular (B) lattice for the obstacle set $\Lambda$ does not affect the structure of the tangling phase diagram.
    (C)~Changing the $R$-axis reveals a range of $R$-values for which tangling is mostly controlled by $\gamma$, with small $\gamma$ producing untangled states and large $\gamma$ producing tangled states. All phase diagrams (A-C) have $D_R/\alpha = 0.4,\, r/\ell=1/3,\, L/\ell=25$.
    (D-F)~Sample trajectories run for $0<t<0.4L$ corresponding to points on the phase diagrams (top row) marked by black circles. Disks indicate obstacles in $\Lambda$, colored by the value of their contact winding, $cW_p$, due to the displayed trajectory. (D,E) Near the critical tangling index, $\mathcal{T}^\ast \approx 2$, trajectories typically wind around points in both the clockwise and counterclockwise directions. This can cause winding cancellation, where a point has low winding despite being enclosed by a curve (D). These trajectories could be part of a topological quick release mechanism, and are reminiscent of the ``picture-hanging puzzle"~\cite{demaine2014picture}.
    (F)~Trajectories with large $R$ have very small curvature, thus leading to untangled states. Friction effects often prevent these linear motions from being a feasible untangling method. Trajectories in (D-F) have low noise ($D_R/\alpha = 0.2$) in order to illustrate their topological properties.}
	\label{SI_mean_field_curves}
\end{figure*}

The mean-field trajectories described by our phase diagrams (Fig.~\ref{SI_mean_field_curves}A-C) capture a range of tangling and untangling gaits, including near-linear motion at large $R$ (Fig.~\ref{SI_mean_field_curves}C). In particular these gaits often display interesting topological motifs. For example, trajectories with tangling index close to critical ($\mathcal{T}^\ast\approx 2$) typically wind around points in both the clockwise and counterclockwise directions (Fig.~\ref{SI_mean_field_curves}D,E). This chirality switching behavior can cause the trajectory to cancel out an earlier winding by passing around the same point twice in opposite directions (Fig.~\ref{SI_mean_field_curves}D). Physically if a worm winds clockwise around a point $p$, then clockwise around another point $p'$, and finally anticlockwise around $p$ once more, the worm will still be entangled with $p$ due to winding around $p'$. These multi-worm interactions are not captured by our mean-field model, however, the presence of such winding cancellation effects in critical worm trajectories suggests that critical tangles are equipped with additional topological quick release mechanisms for rapid untangling. We note that the trajectories in Fig.~\ref{SI_mean_field_curves}D,E resemble the solution to the famous ``picture-hanging puzzle"~\cite{demaine2014picture}, which is an example of one such quick release mechanism.

Expanding the $R$-axis of our phase diagrams demonstrates that extreme values of $R$ produce untangling gaits (Fig.~\ref{SI_mean_field_curves}C, F). However, the small $R$ untangling gait requires producing very high curvatures, and is therefore energetically unfavorable. Similarly, the large $R$ trajectory appears nearly linear (Fig.~\ref{SI_mean_field_curves}F). However attempting linear motion from an initially tangled state will incur a high friction cost, owing to capstan-like friction effects~\cite{sano2022exploring}. Elastic relaxation to such a linear state will also require relatively long timescales.

\subsection*{Estimation of model parameters from data}

\begin{figure*}[t]
	\centering
	\includegraphics[width=\textwidth]{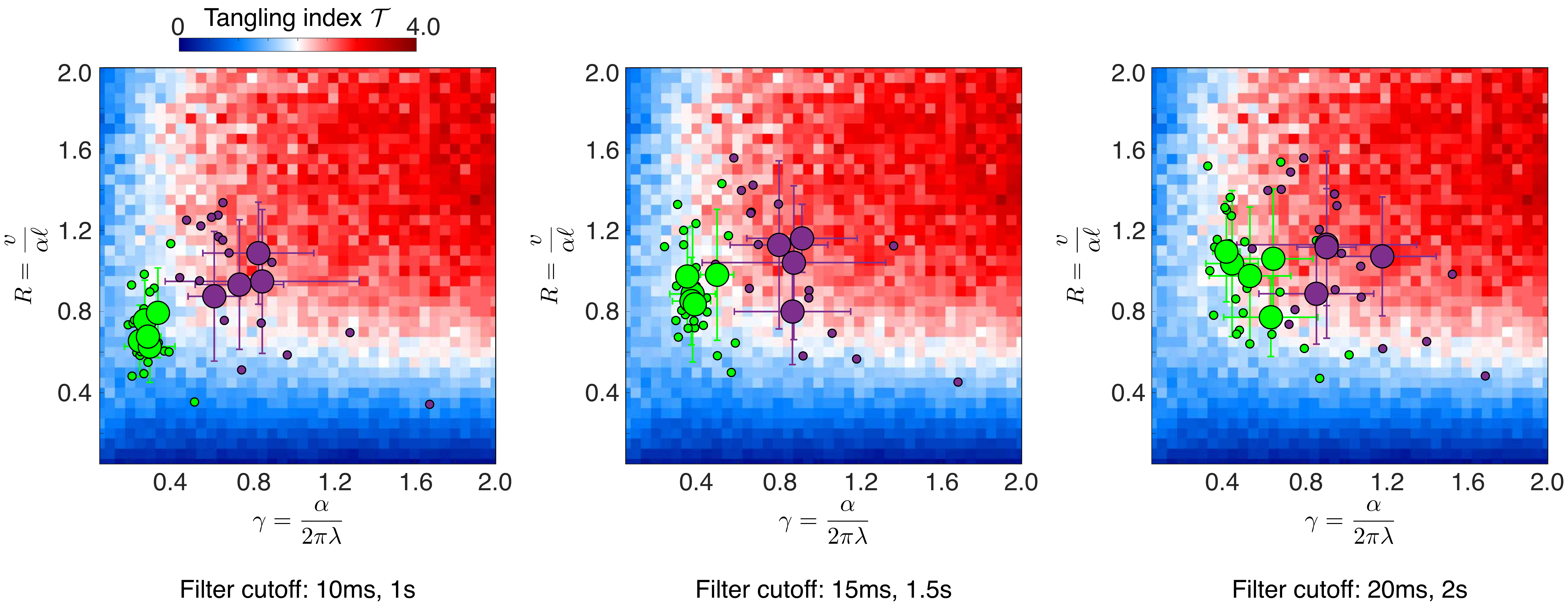}
	\caption{\textbf{Effect of smoothing on estimated worm head trajectory parameters}
	Worm head trajectory data are smoothed using a lowpass filter. The estimated parameters depend only weakly on the choice of cutoff frequency. In Fig.~4C we choose frequency cutoffs of $15$ms and $1.5$s (middle diagram). Background phase diagrams have $D_R/\alpha = 0.4,\, r/\ell=1/3,\, L/\ell=25$ (Fig.~\ref{SI_phase_diagrams}) as in Fig.~4C.}
	\label{SI_cutoff_frequencies}
\end{figure*}

Here we describe the estimation of the quantities $\alpha,\lambda,v,\ell$. The parameters $\alpha,\lambda, v$ are obtained from trajectory data, and $\ell$, which is a measure of tangle tightness, is obtained from the ultrasound data sets. 

We begin with the procedure for estimating $\alpha, \lambda, v$, by fitting discrete trajectory data $\bs{x}_i = \bs{x}(t_i)$, with $t_{i+1}-t_i = \delta t$, to our SDE model (equation \ref{SDE2})
\begin{align*}
d\mathbf{X} &= v \mathbf{N}_\theta \, dt + \sqrt{2D_T}  * d\mathbf{B}(t)\\
d\mathbf{\theta} &= (-1)^{S(t)} \alpha \, dt + \sqrt{2D_R} * d W(t) 
\end{align*}
As stated previously, we do not expect the noise in the real, biological system to be Gaussian. Instead we proceed by smoothing the trajectory, and we assume that this provides sufficient denoising to estimate $\alpha,\lambda,v$, directly.

We first smooth $\bs{x}(t_i)$ to obtain $\tilde{\bs{x}}_i$. The smoothing operation used was the inbuilt lowpass filter in MATLAB R2021a. The cutoff frequency chosen was $1/\omega = 15\,$ms for the untangling trajectories, and $1/\omega = 1.5\,$s for the tangling trajectories. However the estimated values of $\gamma = \alpha/2\pi\lambda$ and $R = v/\alpha\ell$ do not depend strongly on the choice of cutoff frequency (Fig.~\ref{SI_cutoff_frequencies}). From $\tilde{\bs{x}}$ we reconstruct the velocity and angle 
\begin{align*}
\tilde{\bs{v}}_i &= \frac{\tilde{\bs{x}}_{i+1} - \tilde{\bs{x}}_{i-1}}{2\delta t} \\
\theta_i &= \arg \tilde{\bs{v}}_i
\end{align*}
Finally, let $\tilde{\theta}_i$ be the result of passing $\theta_i$ through the same lowpass filter as $\bs{x}(t)$. We assume that $\tilde{\bs{x}}_i, \tilde{\bs{v}}, \tilde{\theta}_i$ are then smooth enough that $\alpha, \lambda$ and $v$ can be estimated directly.
\begin{align*}
    \alpha = \left\langle \left( \frac{\tilde{\theta}_{i+1} - \tilde{\theta}_i}{\delta t} \right)^2 \right\rangle^{1/2} ,\qquad     v = \left\langle \tilde{\bs{v}}_i \cdot \tilde{\bs{v}_i} \right\rangle^{1/2} 
\end{align*}
This assumption is valid when the noise in the trajectory data is concentrated in the high frequency sector.

The value of $\ell$ will typically depend on how loose or tight the tangle is. We choose $\ell$ from a tangle follows. Let $\bs{x}_i(s)$, denote the centerline curve of the $i$'th worm, where $s\in [0,L_i]$ and $L_i$ is the length of the $i$'th worm. We take $\ell$ to be the average minimum distance between a worm head and the tangle
\begin{align}\label{l_measurer}
    \ell =  \frac{1}{N} \sum_i \min_{s,j\neq i} | \bs{x}_i(L_i) - \bs{x}_j(s)|
\end{align}
where the curves are oriented so $\bs{x}_i(L_i)$ is the worm head, and $N$ is total number of worms. Since the worm heads and tails are indistinguishable in our ultrasound data sets, in this case we measure $\ell$ by
\begin{align*}
    \ell =  \frac{1}{2N} \sum_i \min_{s,j\neq i} | \bs{x}_i(L_i) - \bs{x}_j(s)| + \frac{1}{2N} \sum_i \min_{s,j\neq i} | \bs{x}_i(0) - \bs{x}_j(s)|
\end{align*}
For our ultrasound reconstructions, this calculation gives $\ell\approx 1.5$mm, or $\ell \approx 3h$ in terms of the worm radius.

\subsection*{Tangling and untangling simulations}

\begin{figure*}[t]
	\centering
	\includegraphics[width=\textwidth]{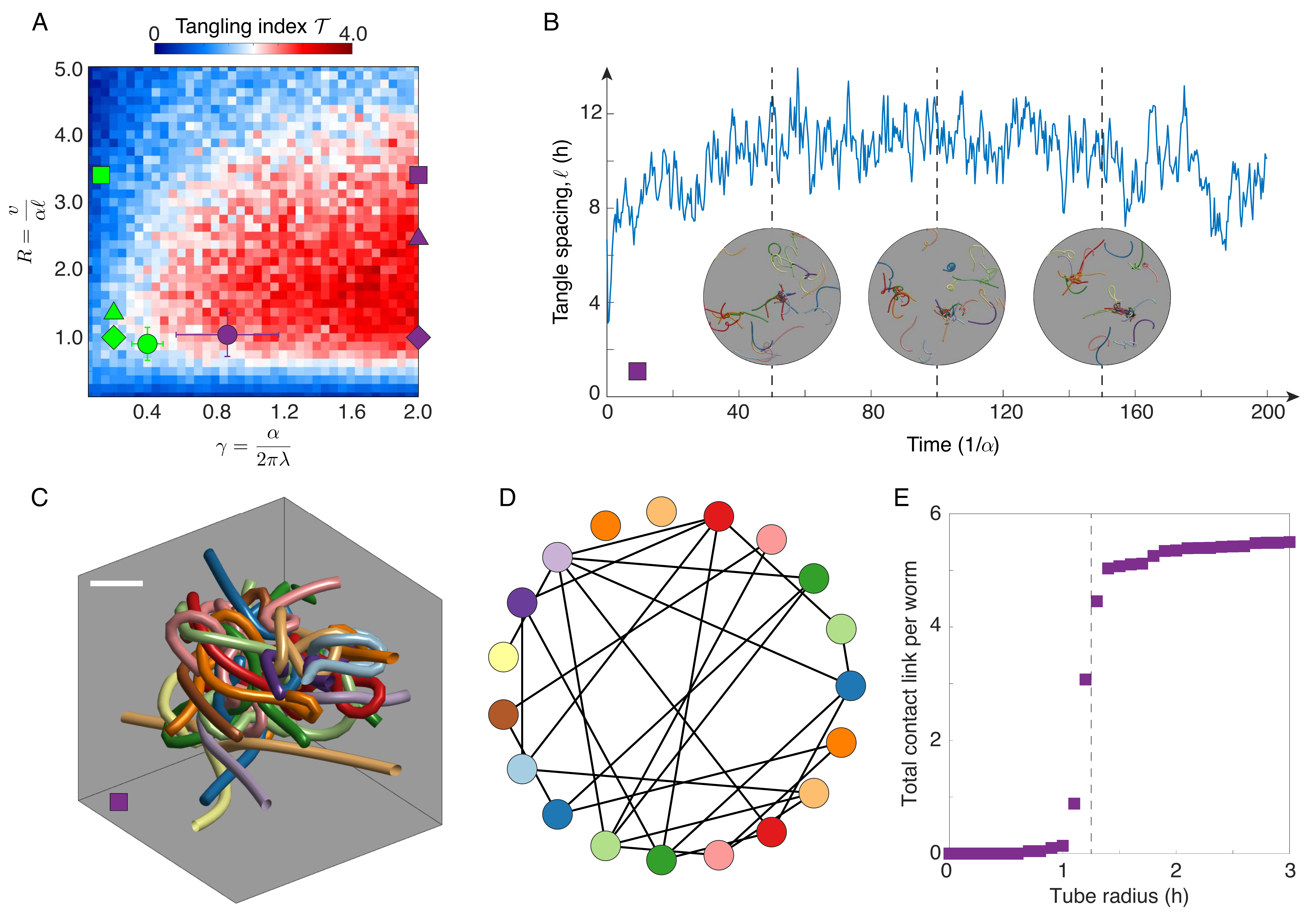}
	\caption{\textbf{Tangling and untangling simulations and their corresponding mean-field parameters}
	(A)~Mean-field parameters corresponding to tangling (purple markers) and untangling (green markers) gaits from experiments (circles) and simulations (triangles, squares and diamonds). Circles represent average values over data from 25 worms (green) and 18 worms (purple) shown in Fig.~4C. Triangles denote simulations from Fig.~3E and movie S2; squares and diamonds denote simulations from Fig.~4D and movie S3. The squares represent simulations with low head speed (Fig.~4D, middle column) and diamonds represent simulations with large head speed (Fig.~4D, right hand column). The mean-field phase diagram is predictive of whether the simulations result in tangled or untangled states (movies S2 and S3). 
	(B)~The tangle spacing $\ell$ fluctuates during a tangling simulation (Fig.~4D and movie S3). We measure $\ell$ by taking the average value between $t=50/\alpha$ and $t=100/\alpha$. Mean-field simulation parameters given by purple square in (A). 
	(C)~$19$-worm artificial tangle obtained by taking the largest cluster of touching worms generated by the simulation analyzed in (B) and shown in Fig.~4D (middle column) and movie S3. This tangle is used as the initial condition for the large speed simulations show in Fig.~4D (right hand column) and movie S3. Scale bar~10$h$.
	(D)~Tangle graph for the artificial tangle in (C). Edges are present between worms which have contact link greater than $1/2$. The tangle graph is not connected since the artificial tangle in (C) is chosen as a cluster of touching, but not necessarily tangled, worms.
	(E)~The relationship between total contact link per worm and tube radius for this artificial tangle is similar to those observed for real living tangles (Fig.~2F). The tube radius dependence of total contact link also identifies an effective radius worm radius, $h_{\text{eff}} = 1.25 h$ that arises from contact handling in simulations.
	 }
	\label{SI_simulated_tangles}
\end{figure*}

Our tangling and untangling simulations (Figs.~3-4, movies~S2-S3) agree with mean-field predictions (Fig.~\ref{SI_simulated_tangles}A). These simulations use the elasticity framework and the stochastic active head force \eqref{active_force} discussed above. The parameters $\alpha, \lambda, v$ can therefore be directly input. On the other hand, $\ell$ must be measured. For simulations which begin from a tangled initial condition, a value of $\ell$ can be found by applying \eqref{l_measurer} to this initial condition. On the other hand, during a tangling simulation, the initial condition may not reflect the equilibrium value of $\ell$ (Fig.~\ref{SI_simulated_tangles}B). To measure the tangle tightness in this case, we take the average value, $\langle \ell\rangle$, between $t = 50/\alpha$ and $t = 100/\alpha$. Empirically, this appears to give the value of $\ell$ time to equilibrate from the initial condition, without including the decrease in $\ell$ that occurs once tangled clusters begin to form (Fig.~\ref{SI_simulated_tangles}B). 

In our simulations, we find values of $\ell$ that range from $7h$ to $10h$, where $h$ is the worm radius. In contrast, our ultrasound reconstructions have $\ell \approx 3h$. This discrepancy could arise from a variety of factors. For example the real worms are soft and compressible, which could cause tighter packings. The worms in the ultrasound data sets also have a higher variance in length (Fig.~\ref{SI_ultrasounds}) and smaller average length ($24\,$mm) than the simulated worms (constant length $40\,$mm). The range of different lengths could enable a collective to pack more densely. Finally, we note that the ratio $\ell/h$ could be sensitive to the precise contact handling method used in simulations, and so part of the observed difference in the value of $\ell/h$ could be due to the discretization error.

Overall, generated tangles have similarities to ultrasound reconstructions (Fig.~\ref{SI_simulated_tangles}C-E). In particular, the radius dependence of total contact link per worm exhibits a similar shape for the generated tangle (Fig.~\ref{SI_simulated_tangles}E) as for the ultrasound data sets (Fig.~2F). The value of total contact link per worm in the tangle shown in Fig.~\ref{SI_simulated_tangles}C is 3.6, which is larger than the observed values for ultrasound tangles (1.7 to 2.1, Fig.~2F). We attribute this to the fact that the dynamics used to generate this artificial tangle correspond to a larger tangling index (Fig.~\ref{SI_simulated_tangles}A) than is observed for tangling worms (Fig.~4C).

\subsubsection{Simulation parameters}

Our simulation parameters correspond to the following values for worm radius $h$, worm length $L$, density $\rho$ and bending modulus $E_b$
\begin{align*}
h = 0.5\,\text{mm} , \quad L = 40\,\text{mm}, \quad \rho = 10^{-3}\,\text{g mm}^{-3} , \quad E_b = 1 \,\text{kPa}
\end{align*}
In addition we set an effective bulk modulus $K$, damping coefficients $\gamma, \eta$, friction $\zeta$ and effective Young's modulus~$E$. The first three parameters are fixed across simulations
\begin{align*}
K = 170\,\text{kPa} , \quad \gamma = 0.1\,\text{g mm}^{-3}\text{s}^{-1} ,\quad \eta = 3.7\,\text{g mm}^{-1}\text{s}^{-1}
\end{align*}

Our simulations reproduce tangling and untangling behavior (Fig.~\ref{SI_simulated_tangles}A) for a range of values of the effective Young's modulus~$E$. The tangling and untangling simulations in Fig.~3E (movie S2) and the low head speed simulations in Fig.~4D (movie S3) have $E = 10\,$kPa and the large head speed simulations in Fig.~4D (movie S3) have $E = 100\,$kPa. Note that the order of magnitude of the values chosen for $E_b, E$ and $K$ all fall within the observed range of elastic moduli for \textit{C. elegans}~\cite{gilpin2015worms}. Finally, our simulations are also robust to changes in the friction parameter $\zeta$. For the tangling simulation in Fig.~3E and the low head speed simulations in Fig.~4D, we take $\zeta = 2.7\,\text{g mm}^{-2}\text{s}^{-1}$. The untangling simulation in Fig.~3E has $\zeta = 2.2\,\text{g mm}^{-2}\text{s}^{-1}$ and the large head speed simulations in Fig.~4D have $\zeta = 2.0\,\text{g mm}^{-2}\text{s}^{-1}$.

The parameters of the active head forcing (equation \ref{active_force}) are indicated in Fig.~\ref{SI_simulated_tangles}A, and given explicitly below. The force noise term is constant across all simulations, $D_F = 0.3\, \text{g}^2\text{mm}^{2}\text{s}^{-3}$. The rotational noise strength varies from $D_R = 5\,\text{s}^{-1}$ to $D_R = 10\,\text{s}^{-1}$ across simulations, which corresponds to $D_R/\alpha = 0.6$ for tangling simulations and $D_R/\alpha = 0.04$ for untangling simulations. The other parameters (to 2.s.f) are
\begin{align*}
    v = 110\,\text{mm s}^{-1} , \quad \ell = 5.0\,\text{mm} , \quad \alpha = 8.4\,\text{s}^{-1} , \quad \lambda = 0.7 \,\text{s}^{-1} \qquad &\text{(Fig.~\ref{SI_simulated_tangles}A, purple triangle)} \\
    v = 410\,\text{mm s}^{-1} , \quad \ell = 1.7\,\text{mm} , \quad \alpha = 170\,\text{s}^{-1} , \quad \lambda = 130 \,\text{s}^{-1} \qquad &\text{(Fig.~\ref{SI_simulated_tangles}A, green triangle)} \\
    v = 160\,\text{mm s}^{-1} , \quad \ell = 5.5\,\text{mm} , \quad \alpha = 8.4\,\text{s}^{-1} , \quad \lambda = 0.7 \,\text{s}^{-1} \qquad &\text{(Fig.~\ref{SI_simulated_tangles}A, purple square)} \\
    v = 160\,\text{mm s}^{-1} , \quad \ell = 5.2\,\text{mm} , \quad \alpha = 8.4\,\text{s}^{-1} , \quad \lambda = 11 \,\text{s}^{-1} \qquad &\text{(Fig.~\ref{SI_simulated_tangles}A, green square)} \\
    v = 900\,\text{mm s}^{-1} , \quad \ell = 3.3\,\text{mm} , \quad \alpha = 250\,\text{s}^{-1} , \quad \lambda = 20 \,\text{s}^{-1} \qquad &\text{(Fig.~\ref{SI_simulated_tangles}A, purple diamond)} \\
    v = 900\,\text{mm s}^{-1} , \quad \ell = 3.3\,\text{mm} , \quad \alpha = 250\,\text{s}^{-1} , \quad \lambda = 200 \,\text{s}^{-1} \qquad &\text{(Fig.~\ref{SI_simulated_tangles}A, green diamond)}  
\end{align*}
Our phase diagram therefore predicts the outcome of multi-filament simulations over a wide range of parameters, including for different values of noise.

\bibliography{references}